\documentclass[aps,pra,twocolumn,groupedaddress,showpacs]{revtex4}
\usepackage{bm}
\usepackage{epsf}
\usepackage{amssymb}
\usepackage{amsmath}
\usepackage{graphicx}
\usepackage{rotating}
\usepackage{epsfig}
\usepackage{psfrag}
\usepackage{amsmath}
\usepackage{hyperref}

\hypersetup{
    bookmarks=true,         
    unicode=false,          
    pdftoolbar=true,        
    pdfmenubar=true,        
    pdffitwindow=true,      
    pdftitle={My title},    
    pdfauthor={Author},     
    pdfsubject={Subject},   
    pdfcreator={Creator},   
    pdfproducer={Producer}, 
    pdfkeywords={keywords}, 
    pdfnewwindow=true,      
    colorlinks=false,       
    linkcolor=red,          
    citecolor=green,        
    filecolor=magenta,      
    urlcolor=cyan           
}

\DeclareMathAlphabet{\bi}{OML}{cmm}{b}{it}

\begin{document}
\def\G{{\cal G}}
\def\F{{\cal F}}
\def\ea{\textit{et al.}}
\def\bM{{\bm M}}
\def\bN{{\bm N}}
\def\bV{{\bm V}}
\def\bj{\bm{j}}
\def\bSig{{\bm \Sigma}}
\def\bLam{{\bm \Lambda}}
\def\bfeta{{\bf \eta}}
\def\bn{{\bf n}}
\def\d{{\bf d}}
\def \xy{$x$--$y$ }
\def\bP{{\bf P}}
\def\bK{{\bf K}}
\def\bk{{\bf k}}
\def\bkn{{\bf k}_{0}}
\def\bx{{\bf x}}
\def\bz{{\bf z}}
\def\bR{{\bf R}}
\def\br{{\bf r}}
\def\bu{{\bm u}}
\def\bq{{\bf q}}
\def\bp{{\bf p}}
\def\by{{\bf y}}
\def\bQ{{\bf Q}}
\def\bs{{\bf s}}
\def\bA{{\mathbf A}}
\def\bv{{\bf v}}
\def\b0{{\bf 0}}
\def\la{\langle}
\def\ra{\rangle}
\def\Im{\mathrm {Im}\;}
\def\Re{\mathrm {Re}\;}
\def\beq{\begin{equation}}
\def\eeq{\end{equation}}
\def\bea{\begin{eqnarray}}
\def\eea{\end{eqnarray}}
\def\bdm{\begin{displaymath}}
\def\edm{\end{displaymath}}
\def\bnab{{\bm \nabla}}
\def\Tr{{\mathrm{Tr}}}
\def\bJ{{\bf J}}
\def\bU{{\bf U}}
\def\bPsi{{\bm \delta\Delta}}
\def\mA {\mathrm{A}}
\def \R{R_{\mathrm{s}}}
\def \rhos{n_{\mathrm{s}}}
\def \rhon{\tilde{n}}
\def \Rd{R_{\mathrm{d}}}
\def \xy{three dimensional $XY\;$}
\def\sfrac{\textstyle\frac}
\def \leq{\mathrm{l.e.}}

\title{Viscosity of strongly interacting quantum fluids: spectral functions and sum rules}
\author{Edward~Taylor}
\affiliation{Department of Physics, The Ohio State University, Columbus, Ohio, 43210}
\author{Mohit~Randeria}
\affiliation{Department of Physics, The Ohio State University, Columbus, Ohio, 43210}

\date{August 20, 2010}

\begin{abstract}
The viscosity of strongly interacting systems is a topic of great interest in diverse fields.
We focus here on the bulk and shear viscosities of \emph{non-relativistic} quantum fluids, with particular emphasis on 
strongly interacting ultracold Fermi gases. We use Kubo formulas for the bulk and shear viscosity spectral functions,
$\zeta(\omega)$ and $\eta(\omega)$ respectively, to derive exact, non-perturbative results.
Our results include: a microscopic connection between the shear viscosity $\eta$ and the normal fluid density $\rho_n$;
sum rules for $\zeta(\omega)$ and $\eta(\omega)$ and their evolution through the BCS-BEC crossover; 
universal high-frequency tails for $\eta(\omega)$ and the dynamic structure factor $S({\bf q}, \omega)$.
We use our sum rules to show that, at unitarity, $\zeta(\omega)$ is identically zero and thus relate 
$\eta(\omega)$ to density-density correlations. We predict that frequency-dependent shear viscosity
$\eta(\omega)$ of the unitary Fermi gas can be experimentally measured using Bragg spectroscopy.
\end{abstract}
\maketitle

\section{Introduction}

The study of the viscosity of strongly interacting quantum fluids has brought together very different areas of
physics -- black holes and string theory, quark-gluon plasmas, quantum fluids and cold atoms -- which, at first sight,
appear to have little in common~\cite{Son-review,Schafer09}. 
This extraordinary development originated with the work of Son, Starinets and coworkers~\cite{Son-review,Policastro01,Kovtun05} 
who calculated the shear viscosity in a strongly interacting quantum field theory,
the ${\cal{N}}=4$ supersymmetric Yang-Mills (SYM) theory,
and conjectured a lower bound 
\beq
{\eta}/{s} \ge {\hbar}/{(4\pi k_B)}
\label{bound}
\eeq
for the ratio of the shear viscosity $\eta$ to the entropy density $s$ of \emph{any} system. 
These results were obtained using the AdS/CFT formalism where certain strongly coupled 
field theories can be mapped onto weakly coupled gravity theories.

Although a number of counterexamples have since been proposed~\cite{Cohen07,Brigante08,Kats09,Buchel09}, there are no known experimental violations of the bound given by Eq.~(\ref{bound}).  Remarkably, two very different experimental systems come close to saturating the bound:
the quark-gluon plasma at Brookhaven's Relativistic Heavy Ion Collider~\cite{RHIC,Teaney01}, and  
ultracold atomic Fermi gases~\cite{Turlapov08,Gelman05} 
close to a Feshbach scattering resonance, where the $s$-wave scattering length becomes infinite~\cite{Trentoreview}.
This is the strongly interacting unitary regime that lies at the center of the BCS-BEC crossover.  These two systems are amongst the hottest and coldest systems every realized in a laboratory.  

\medskip
In this paper we focus on \emph{non-relativistic} quantum fluids, with particular emphasis on \emph{strongly
interacting Fermi gases}.  These are systems for which the most controlled experiments should be possible. 
A ``perfect fluid'' with the minimum shear viscosity is necessarily in a quantum regime, since the bound involves $\hbar$.
In addition, it must also be in a strongly interacting regime where \emph{well-defined quasiparticle excitations do not exist}. 
If the system had sharp quasiparticles, then their mean scattering rate $\tau^{-1}$ would be much less than the average energy
per particle $\epsilon_0$, so that $\hbar/\tau \ll \epsilon_0$. We can then use Boltzmann's kinetic theory approach to obtain
$\eta \sim n \epsilon_0 \tau$, where $n$ is the number density. 
Using $s \sim nk_B$, we find a large $\eta/s \sim \epsilon_0 \tau / k_B \gg {\hbar}/{k_B}$.
Thus, in order to find perfect fluids that come close to saturating the lower bound given by Eq.~(\ref{bound}), one must
look at strongly interacting quantum fluids where the quasiparticle approximation fails.

In this paper we use Kubo formulas for the frequency-dependent
spectral functions for shear viscosity $\eta(\omega)$ and
bulk (or second) viscosity $\zeta(\omega)$, and derive several exact, non-perturbative results without making 
weak coupling or quasiparticle approximations. Our main results are:

\medskip
$\bullet$ We establish a microscopic connection between the shear viscosity $\eta$ and the normal fluid density $\rho_n$
and show that a non-zero $\rho_n$ is a necessary condition for a non-vanishing $\eta$.

\medskip
$\bullet$ We derive sum rules for $\eta(\omega)$ and $\zeta(\omega)$ of any Bose or Fermi system with 
an arbitrary two-body interaction; see Eqs.~(\ref{etasumruleiso}) and (\ref{zetasumruleiso}).

\medskip
$\bullet$ For a dilute two-component Fermi gas, we find the shear viscosity sum rule
\bea 
\frac{1}{\pi}\int^{\infty}_{0}\!\!d\omega \left[\eta(\omega) -\frac{C}{10\pi\sqrt{m\omega}}\right]=
\frac{\varepsilon}{3}-\frac{ C}{10\pi m a},
\label{etasumrule0finite}
\eea
valid for arbitrary temperature and $1/(k_F a)$, where $a$ is the $s$-wave scattering length.
Here, $\varepsilon$ is the energy density and $C$ is the \textit{contact}~\cite{Tan08}.  A central quantity in many of our results, 
$C = k_F^4 {\cal C}[1/(k_F a), T/\epsilon_F]$ can be defined via the large-$k$ tail of the momentum distribution 
$n_{\bk} \simeq C/k^4$ for $k \gg k_F$, and characterizes the short-distance properties of the many-body state.  

\medskip
$\bullet$ For the bulk viscosity, we obtain the sum rule
\bea 
\frac{1}{\pi}\int^{\infty}_{0}d\omega\; \zeta(\omega)=\frac{1}{72 \pi m a^2} \left( \frac{\partial C}{\partial a^{-1}} \right)_s,
\label{zetasumrule0}
\eea
where the derivative is at fixed entropy per particle $s \equiv S/N$. [Different from Eq.~(\ref{bound}), in the remainder of this paper we use $s$ to denote this quantity rather than the entropy density.] 
Below the superfluid transition, the bulk viscosity that enters Eq.~(\ref{zetasumrule0}) is $\zeta_2$, 
associated with the damping of in-phase motions of the superfluid and normal components. The positivity
of $\zeta(\omega)$ implies that $(\partial C / \partial a^{-1})_s \geq 0$.

\medskip
$\bullet$ At unitarity, the bulk viscosity spectral function vanishes at \emph{all} frequencies and \emph{all} temperatures. Quite generally, $\zeta(\omega) \geq 0$,
but the sum rule in Eq.~(\ref{zetasumrule0}) vanishes for $|a|=\infty$ and thus $\zeta(\omega) = 0$ for the unitary Fermi gas. 
This generalizes the result~\cite{Son07} that the static bulk viscosity $\zeta(0)$ vanishes at unitarity.

\medskip
$\bullet$ It follows from the previous result that, at unitarity, the shear viscosity spectral function $\eta(\omega)$ can be
related to density-density correlations as
\bea 
\eta(\omega)
= \lim_{q\to 0}\frac{3\omega^3}{4 q^4}\;\mathrm{Im}\chi_{\rho\rho}(\bq,\omega) \ \ \ \ \ (|a| = \infty).
\eea
Thus, $\eta(\omega)$ for the unitary Fermi gas
can be measured spectroscopically using, for instance, two-photon Bragg spectroscopy.

\medskip
$\bullet$ We show from our sum rules that various dynamic response functions for Fermi gases
have high-frequency tails characterized by odd-integer power laws,
whose magnitudes are controlled by the contact $C$. The tail $C/\sqrt{\omega}$  of $\eta(\omega)$ is evident from 
Eq.~(\ref{etasumrule0finite}). Using this we find that the dynamic structure factor has a tail of the form~\cite{Son10}
 $S({\bq},\omega) \sim C q^4/\omega^{7/2}$ for
$q \to 0$ and $\omega \to \infty$, which is shown to be a generic feature of short range physics. 

\bigskip
In the remainder of this Section we describe how the rest of the paper is organized.
In Section~\ref{Kubosec}, we begin with a careful derivation of Kubo formulas for the spectral functions
$\eta(\omega)$ and $\zeta(\omega)$ in terms of current-current correlation functions,  
Eqs.~(\ref{KuboTR}) and (\ref{KuboLR}), and, equivalently, in terms of the stress-stress correlator, Eq.~(\ref{KuboPiFull}).

In Section~\ref{rhon-section} we recall some elementary facts about the shear viscosity of a fluid and why it
is analogous to the resistivity, and not the conductivity, of a metal. We then establish a connection between the viscosity $\eta$ and 
the normal fluid density $\rho_n$ using microscopic response functions.

After establishing the positivity of $\eta(\omega)$ and of $\zeta(\omega)$ in Section~\ref{posdefsec},
we derive sum rules for these quantities in Section~\ref{sumrulesec}. The most general sum rules for
the shear and bulk viscosities of any Bose or Fermi system with an arbitrary isotropic interaction potential $V(p)$,
and valid for all temperatures, are given in Eqs.~(\ref{etasumruleiso}) and (\ref{zetasumruleiso}).

In Section~\ref{diluteviscositysec}, we specialize to the dilute Fermi gas, with interparticle spacing
$k_F^{-1}$ and $s$-wave scattering length $a$ both much larger than the characteristic range $r_0$ of the potential.
We obtain the $\zeta$ sum rule in Eq.~(\ref{zetasumrule0}), which is finite in the zero range limit.
The $\eta$ sum rule, however, has an ultraviolet divergence; see Eq.~(\ref{etasumrule2}).
We identify, in Section~\ref{highfrequencytailsec}, the $C/\sqrt{\omega}$ high-frequency tail of the shear viscosity spectral function,
and derive the sum rule given by Eq.~(\ref{etasumrule0finite}), which is manifestly finite for $r_0 \to 0$. 
The sum rules given by Eqs.~(\ref{etasumrule0finite}) and (\ref{zetasumrule0}) are valid in both the normal and superfluid phases,
with $\zeta$ replaced by $\zeta_2$ in the latter state.

In Section~\ref{Crossoversec} we show from the $\zeta$ sum rule that, at unitarity, $\zeta(\omega)$ vanishes at all frequencies
and all temperatures.
We also discuss the $1/(k_F a)$-dependence of the $\eta$  and $\zeta$ sum rules across the BCS-BEC crossover, using
available quantum Monte Carlo data for the energy density at $T=0$.

We discuss the connection between viscosity and density-density correlations in Section~\ref{SqomegaSec} and find
two interesting results. First, we show how a density probe such as two-photon Bragg spectroscopy can in principle be used to measure the shear
viscosity spectral function $\eta(\omega)$ at unitarity.
Second, we identify the high-frequency $\omega^{-7/2}$ tail in the dynamic structure factor $S({\bq},\omega)$.
 
In Section~\ref{comparisonsec}, we briefly compare the sum rules that we have derived for non-relativistic
quantum fluids with those obtained in relativistic quantum field theories. Finally in Section~\ref{conclusionssec}
we conclude with open questions.

There are five Appendices which contain technical details of derivations or review certain results which are used at
various places in the paper. In Appendix~\ref{KuboPi0}, we briefly discuss an alternate stress tensor operator often used to calculate the shear viscosity. Some results from dissipative two-fluid hydrodynamics, which we use in our paper, are reviewed in Appendix~\ref{hydrosec}.  
We review in Appendix \ref{ContactAppendix} results related to the contact that are used at several places in the paper, 
and also give a detailed derivation of certain equations that involve the contact.   
In Appendix~\ref{Pressuresec}, we derive a microscopic expression for the pressure.  Finally, in
Appendix \ref{universalthermosec} we give details of the derivation of the $\zeta$ sum rule which make use of the scaling form
of thermodynamic functions across the BCS-BEC crossover. 


\section{Kubo formula for viscosity}  
\label{Kubosec}    

We begin by deriving Kubo formulas for the bulk and shear viscosity. Although the results of this Section
are, for the most part, ``well known'', we could not find a complete derivation at any one place in the literature.  
In particular, there are several subtle points not dealt with adequately 
elsewhere, not least the definition of the stress tensor operator
$\widehat{\Pi}_{\alpha\beta}$ for non-relativistic systems. 

To introduce notation, we start with the Euler equation
\bea 
m\partial_t j_{\alpha}(\br,t)= -\partial_{\beta}\Pi_{\alpha\beta}(\br,t),
\label{euler}
\eea
where $m$ is the mass of the particles, $j_{\alpha}$ is the (number) current and 
$\Pi_{\alpha\beta}$ is the momentum flux density tensor, which we call the \emph{stress tensor}, for short. 
Here, $\alpha$ and $\beta$ take on values $x,y,z$ (and there is no difference between upper and lower indices 
in our non-relativistic formulation).
In general, the stress tensor is given by~\cite{LLFM}
\bea 
\Pi_{\alpha\beta} = P\delta_{\alpha\beta} + \rho u_{\alpha}u_{\beta} -\sigma'_{\alpha\beta},
\label{stress}
\eea
where $P$ is the pressure, $\rho$ the mass density and $\bu$ the velocity. 
The viscous term $\sigma'_{\alpha\beta}$ is given by
\bea 
\sigma'_{\alpha\beta}=  \eta\left[\partial_{\beta} u_{\alpha} \!+\! \partial_{\alpha}u_{\beta} \!-\! \frac{2}{3}\delta_{\alpha\beta}(\bnab\cdot\bu)\right] \!+\! \zeta\delta_{\alpha\beta}(\bnab\cdot\bu),
\label{stresstensor}
\eea 
where $\eta$ is the shear viscosity and $\zeta$ the bulk viscosity.
The generalization of Eq.~(\ref{stresstensor}) to the superfluid state is well known~\cite{LLFM} and
involves additional bulk viscosities. 
At the end of Section~\ref{currentcorrelatorsec}, we show that the Kubo formula we derive for $\zeta$ 
describes the bulk viscosity $\zeta_2$ in the superfluid phase.  

Our goal is to obtain Kubo formulas for frequency-dependent generalizations of 
the long-wavelength viscosities, $\eta$ and $\zeta$, in terms of equilibrium correlation functions
of the many-body system.
The Kubo formulas for viscosities are often written in terms of the stress-stress
correlators; see, e.g., Sec.~90 of Ref.~\cite{LLStatPhysII}. 
However, the form of the stress tensor (or momentum flux density) \emph{operator} 
$\widehat{\Pi}_{\alpha\beta}$ is not obvious, and many different, complicated
expressions \cite{Forster} which are presumably equivalent can be found in the literature.
Part of the problem is to write down an operator expression 
for the pressure $P$ in terms of particle positions and momenta.  In high-energy physics, a simple way to calculate the stress-energy tensor $\widehat{\Pi}_{\alpha\beta}$ is to vary the action with respect to the metric in curved space-time.  We prefer, however, to describe non-relativistic fluids without going to curved space-time.

To begin with, in~\ref{currentcorrelatorsec}, we adopt an approach that permits us to get around the 
complexities of defining the stress operator $\widehat{\Pi}$. We 
consider the linear response of a fluid to an externally imposed velocity field
and derive Kubo formulas for the bulk and shear viscosities in terms of \emph{current-current correlators}.  
The results of this subsection are the same as those of Kadanoff and Martin~\cite{Kadanoff63}.

In~\ref{Kubocomparisonsec}, we use an operator form of Euler's equation to make the connection between bulk and shear
viscosities and \emph{stress-stress correlators}.
In Appendix~\ref{KuboPi0}, we derive an alternative form of the stress correlator, which works 
only for the shear viscosity in the zero-frequency limit, but is often used in calculations.

\subsection{Current correlators}
\label{currentcorrelatorsec}

We calculate within linear response theory~\cite{NozieresPines1,Baymbook} the current flow in a fluid 
subjected to an external velocity field $\bu(\br,t) = \bu(\br) e^{-i\omega t} e^{0^+t}$ which is
turned on adiabatically.
Our goal is to relate the imaginary part of this response function 
to viscosity through the dissipative part of the stress tensor.  

The response of a fluid to the ``moving walls'' of its container is a standard concept in the theory of 
superfluidity~\cite{Baymbook}. Here, we generalize this analysis to a non-uniform and time-varying
external perturbation $\bu(\br,t)$, taking the long wavelength limit at the end.
We write the Hamiltonian of the system $\hat{H}$ plus external perturbation $\hat{H}'$ as~\cite{wallnote}
\bea 
\hat{H}_{\rm total} &=& \frac{1}{2m}\sum_{i=1}^N \int d\br\;
\left[\hat{\bp}_i - m\bu(\br,t)\delta(\br-\hat{\br}_i)\right]^2 + \hat{V}\nonumber\\
=\hat{H} \! &-& \!\frac{1}{2}\sum_{i=1}^N \int d\br \bu(\br,t)\cdot
\left\{ \hat{\bp}_i,\delta(\br\!-\!\hat{\br}_i) \right\}
+ {\cal{O}}[\bu^2],\label{Hshift}\eea
where $\hat{\bp}_i$ and $\hat{\br}_i$ are the momentum and position operators, respectively, for the $i$th particle,
$m$ is the mass, and $\hat{V}$ is the potential energy operator. 
The anticommutator $\{ \hat{A},\hat{B} \} = \hat{A}\hat{B} + \hat{B}\hat{A}$ is used to
symmetrize products.
We thus see that to linear order in $\bu$, the external perturbation is
\bea 
\hat{H}'(t) = - m\int d\br e^{-i\omega t} e^{0^+t} \bu(\br)\cdot\hat{\bj}(\br,t),
\label{Hpert}
\eea
where 
$\hat{\bj} = \sum_{i=1}^N \left\{ \hat{\bp}_i , \delta(\br\!-\!\hat{\br}_i) \right\}/2m$
is the current density operator.

Linear response theory gives the result~\cite{NozieresPines1,Baymbook}
\bea 
\lefteqn{\langle \hat{j}_{\alpha}(\br,t)\rangle = m \int d\br'\!\int^{\infty}_{-\infty}\!\!\! dt'e^{0^+t'}e^{-i\omega t'}\times}&&\nonumber\\&&\;\;\;\;\;\;\;\;\;\;\;\; \chi^{\alpha\beta}_{J}(\br-\br',t-t')u_{\beta}(\br').\label{lr1}
\eea
Here and below, we use the standard convention of summing over repeated indices. 
The retarded current correlation function
$\chi^{\alpha\beta}_J$ is obtained by using
$\hat{A}=\hat{j}_{\alpha}$ and $\hat{B}=\hat{j}_{\beta}$ in Eq.~(\ref{chiAB}) below.

For later use, we provide a general definition for the \emph{retarded} response function,
or correlator, for operators $\hat{A}$ and $\hat{B}$:
\bea 
\lefteqn{\chi_{A,B}(\br-\br',t-t') \equiv}&&\nonumber\\&& i\Theta(t-t')\langle[\hat{A}(\br,t),\hat{B}^{\dagger}(\br',t')]\rangle.
\label{chiAB} \eea
Here, $\langle \hat{Q}\rangle = {\rm Tr} [\hat{Q} \exp(-\hat{H}/T)]/{\cal{Z}}$ is
the thermal expectation value at temperature $T$ and 
${\cal{Z}} = {\rm Tr}[\exp(-\hat{H}/T)]$ is the partition function.
The step-function $\Theta(t-t')$ enforces causality.
We will use the convention of unit volume $\Omega = 1$ and set $\hbar = k_B = 1$,
unless explicitly stated otherwise.

We find the spectral representation for $\chi_{A,B}$ using the exact eigenstates and eigenvalues of the fully interacting
many-body Hamiltonian $\hat{H}|a\rangle = E_a|a\rangle$, and Fourier transform the result to obtain
\bea 
\lefteqn{\chi_{A,B}(\bq,\omega) = \frac{1}{{\cal{Z}}}\sum_{a,b}e^{-E_a/T}\times}&&\nonumber
\\&&\left[
\frac{\langle a|\hat{B}^{\dagger}_{\bq}|b\rangle\langle b|\hat{A}_{\bq}|a\rangle}{\omega + E_{ba} + i0^+}
- \frac{\langle a|\hat{A}_{\bq}|b\rangle\langle b|\hat{B}^{\dagger}_{\bq}|a\rangle}{\omega - E_{ba} + i0^+} 
\right],
\label{FourierchiAB}
\eea  
where $E_{ba}\equiv E_b-E_a$.  
The quantity of central interest to us in this paper is the imaginary part of $\chi$, given by
\bea 
\lefteqn{\mathrm{Im}\chi_{A,B}(\bq,\omega) = \pi(1-e^{-\omega/T})}&&\nonumber\\&& 
\times\frac{1}{\cal{Z}}\sum_{a,b}e^{-\beta E_a}\langle a|\hat{A}_{\bq}|b\rangle\langle b|\hat{B}^{\dagger}_{\bq}|a\rangle 
\delta(\omega -E_{ba}).
\label{ImchiAB}
\eea

Returning to the problem of interest, we find that the induced current,
obtained by Fourier transforming Eq.~(\ref{lr1}), is
\bea 
\langle \hat{j}^{\alpha}(\bq,\omega)\rangle = m \chi^{\alpha\beta}_{J}(\bq,\omega)u_{\beta}.
\label{lr}
\eea
$\chi^{\alpha\beta}_J$ is given by Eq.~(\ref{FourierchiAB}) with $\hat{A}_{\bq}=\hat{j}^{\alpha}_{\bq}$ and
$\hat{B}^{\dagger}_{\bq} = \hat{j}^{\beta}_{-\bq}$, where
\bea 
\hat{j}^{\alpha}_{\bq} = \frac{1}{2m}\sum_{\bk\sigma} (2k_{\alpha}+q_{\alpha}) \hat{c}^{\dagger}_{\bk\sigma}\hat{c}_{\bk+\bq\sigma}
\label{jdef}\eea
is the current operator with $\sigma$ denoting the different internal states of interest (e.g., spin).   

Next, we need to relate Eq.~(\ref{lr}) to viscosity,
using ``constitutive relations'' between the current and transport coefficients.
For this we use Eqs.~(\ref{stress}) and (\ref{stresstensor}) substituted into Eq.~(\ref{euler}),
where the symbols $j_{\alpha}$ and $\Pi_{\alpha\beta}$,
\emph{without the hats used for operators}, are understood to 
denote expectation values. In the long-wavelength limit, the contributions to the stress tensor coming from viscous terms dominate
over contributions from pressure fluctuations, while the convective term $\partial_{\beta}u_{\alpha}u_{\beta}$ is beyond linear order in velocity.  We thus get
$m\partial_t j^{\alpha}=\zeta\partial_{\alpha}(\bnab\cdot\bu) + \eta\left[\nabla^2 u_{\alpha}+ \partial_{\alpha}(\bnab\cdot\bu)/3\right]$.
Fourier transforming and comparing with Eq.~(\ref{lr}), we obtain
\bea 
\zeta q_{\alpha}q_{\beta}u_{\beta} + \eta\left(\!q^2u_{\alpha}\! +\! \frac{1}{3}q_{\alpha}q_{\beta}u_{\beta}\!\right)\! =\!- i\omega m^2 \chi^{\alpha\beta}_{J}(\bq,\omega)u_{\beta}.\nonumber\\ 
\label{Kubo1}
\eea

We decompose the current correlation function into its
longitudinal ($\chi_L$) and transverse  ($\chi_T$) components:
\bea 
\chi^{\alpha\beta}_{J} = \chi_L\frac{q_{\alpha}q_{\beta}}{q^2} + \chi_T\left(\delta_{\alpha\beta}-\frac{q_{\alpha}q_{\beta}}{q^2}\right)
\label{chiTL}
\eea 
By taking appropriate $q \to 0$ limits \cite{limits} of Eq.~(\ref{Kubo1}) we find
\bea 
\eta(\omega) = \lim_{q\to 0}{(-i\omega)}m^2 \chi_T(\bq,\omega)/{q^2}
\label{KuboT}
\eea
and
\bea 
\zeta(\omega) + {4\eta(\omega)}/{3} = \lim_{q\to 0}{(-i\omega)}m^2 \chi_L(\bq,\omega)/{q^2}.
\label{KuboL}\eea
These expressions define the \emph{complex} shear and bulk viscosities.
We will be interested in the properties and sum rules of
the \emph{spectral functions}:  
\bea 
\mathrm{Re}\; \eta(\omega) = \lim_{q\to 0}{m^2 \omega}\mathrm{Im}\chi_T(\bq,\omega)/{q^2}
\label{KuboTR}
\eea
and 
\bea 
\mathrm{Re}\; \zeta(\omega) + {4\mathrm{Re} \;\eta(\omega)}/{3} = \lim_{q\to 0}
{m^2 \omega}\mathrm{Im}\chi_L(\bq,\omega)/{q^2}.
\label{KuboLR}
\eea
The static viscosities $\eta$ and $\zeta$ introduced in Eq.~(\ref{stresstensor})
are  $\eta \equiv \mathrm{Re}\; \eta(\omega=0)$ and $\zeta \equiv \mathrm{Re}\; \zeta(\omega=0)$.

In closing this subsection, we note that the Kubo formulas for the viscosity derived here and below
are valid in both the normal and superfluid phases, provided we recognize that the bulk viscosity
in the superfluid state refers to $\zeta_2$, which describes damping associated with an in-phase motion of the superfluid and normal fluid components~\cite{LLFM}.  To understand this in more detail, we recall Landau's two-fluid hydrodynamics~\cite{LLFM} for the
superfluid state. In this theory, three bulk viscosities, $\zeta_1$, $\zeta_2$, and $\zeta_3$,  are required 
to describe the dissipation associated with different types of relative motions of the superfluid and normal components.  
The \emph{longitudinal} response does not distinguish between the superfluid and normal components \cite{Baymbook} and 
thus forces the superfluid and normal fluid velocities to be equal: $\bv_s = \bv_n = \bu$.  When both components 
flow with the same velocity, the two-fluid hydrodynamic stress tensor [see Eq.~(140.5) in Ref.~\cite{LLFM}] reduces to the expression in
Eq.~(\ref{stresstensor}), with $\zeta$ replaced by $\zeta_2$, the bulk viscosity associated with the damping of the
in-phase motions of the superfluid and normal fluid components.   
One can also show by direct application of Eq.~(\ref{KuboLR2}) to the two-fluid
hydrodynamic density response function in Eq.~(\ref{chirhorho}) that the left-hand side of Eq.~(\ref{KuboLR2}) is $\zeta_2+4\eta/3$ in the
low-frequency two-fluid hydrodynamic regime.    

\subsection{Stress correlators}
\label{Kubocomparisonsec}

We next derive Kubo formulas equivalent to those derived above but expressed in terms of
the correlators of a suitably defined stress tensor operator $\widehat{\Pi}_{\alpha\beta}$.
These are useful to make connections with the literature~\cite{Kovtun05,Bruun05,Peshier05}.
We will also use these results in connection with the positivity of the bulk viscosity
spectral function and its vanishing for the unitary Fermi gas. 

The $\widehat{\Pi}_{\alpha\beta}$ operator must satisfy 
\bea 
i m [\hat{j}_{\alpha},\hat{H}] = \partial_{\beta}\widehat{\Pi}_{\alpha\beta},
\label{euler-op}
\eea
which is the operator version of the Euler equation, Eq.~(\ref{euler}), and is simply a statement of momentum conservation.
We go to Fourier space and relate matrix elements of the current operator to those of the stress tensor
by sandwiching Eq.~(\ref{euler-op}) between exact many-body eigenstates.
Using the spectral representation in Eq.~(\ref{FourierchiAB}) we can then relate the current correlator
$\chi^{\alpha\beta}_{J}(\bq,\omega)$ to the stress correlator
$\chi^{\alpha\beta,\mu\nu}_{\Pi}(\bq,\omega)$. The latter is defined by choosing
$\hat{A}=\hat{\Pi}^{\alpha\beta}(\bq)$ and $\hat{B}=\hat{\Pi}^{\mu\nu}(-\bq)$ in Eq.~(\ref{FourierchiAB}).

For simplicity we calculate only $\chi^{xx}_{J}$, which will suffice for our purposes.
The final result, after some simple algebra, is 
\bea 
m^2\omega^2\chi^{xx}_{J}(\bq,\omega) &=&q_{\alpha}q_{\beta}\chi^{x\alpha,x\beta}_{\Pi}(\bq,\omega) -
 mq_{\alpha}\langle[\widehat{\Pi}^{x\alpha}_{\bq},\hat{j}^{x}_{-\bq}]\rangle.\nonumber\\ \!\!\!\!\!\!\!
\label{Correlationrelation}
\eea
Note that $\widehat{\Pi}^\prime_{\alpha\beta} = \widehat{\Pi}_{\alpha\beta} + \hat{\Lambda}_{\alpha\beta}$,
with any symmetric tensor $\hat{\Lambda}$ satisfying $\partial_{\beta}\hat{\Lambda}_{\alpha\beta} = 0$, will
also be a solution to the Euler equation, Eq.~(\ref{euler-op}). 
This non-uniqueness in the definition of $\widehat{\Pi}$ does not affect our final results
for the viscosity, related to $\chi^{xx}_{J}$, since a symmetric $\hat{\Lambda}$
with $q_{\beta}\hat{\Lambda}_{\alpha\beta} = 0$ makes no contribution to
Eq.~(\ref{Correlationrelation}).

Using the decomposition given by Eq.~(\ref{chiTL}), and taking the appropriate limits, we find
\bea 
m^2\omega^2\lim_{q\to 0}\frac{\chi_T}{q^2} = \lim_{q\to 0}
\left[ \chi^{xy,xy}_{\Pi} - \frac{m}{q} \langle[\widehat{\Pi}^{xy}_{\bq},\hat{j}^{x}_{-\bq}]\rangle \right],
\label{TPi}
\eea
where we have taken $q_x$ and $q_z$ to zero before $q_y$, and
\bea m^2\omega^2\lim_{q\to 0}\frac{\chi_L}{q^2} = \lim_{q\to 0} \left[ \chi^{xx,xx}_{\Pi}- \frac{m}{q} \langle[\widehat{\Pi}^{xx}_{\bq},\hat{j}^{x}_{-\bq}]\rangle \right],
\label{LPi}
\eea
where we have taken $q_y$ and $q_z$ to zero before $q_x$. We note that the commutators on the right hand sides
of Eqs.~(\ref{TPi}) and (\ref{LPi}) only affect the real parts of $\chi_T$ and $\chi_L$ and not the
spectral functions of interest, shown in the next two equations.

Using the Kubo formulas given by Eqs.~(\ref{KuboTR}) and (\ref{KuboLR})
that were derived above, we find 
\bea 
\mathrm{Re}\; \eta(\omega) = \lim_{q\to 0}{\mathrm{Im}\chi^{xy,xy}_{\Pi}(\bq,\omega)}/{\omega}
\label{KuboPiT}
\eea
and 
\bea 
\mathrm{Re}\; \zeta(\omega) + {4\mathrm{Re} \;\eta(\omega)}/{3}
= \lim_{q\to 0}{\mathrm{Im}\chi^{xx,xx}_{\Pi}(\bq,\omega)}/{\omega}.
\label{KuboPiL}
\eea
In an isotropic system, in the $q \to 0$ limit, the only fourth rank tensor allowed by
symmetry is of the form $A \delta_{\alpha\beta} \delta_{\mu\nu} 
+ B\left( \delta_{\alpha\mu} \delta_{\beta\nu}  + \delta_{\alpha\nu} \delta_{\beta\mu} \right)$.
We can thus combine Eqs.~(\ref{KuboPiT}) and (\ref{KuboPiL}) to write
\bea 
\left[\mathrm{Re}\; \zeta - \frac{2}{3} \mathrm{Re}\; \eta \right] \delta_{\alpha\beta} \delta_{\mu\nu} 
&+& \mathrm{Re}\; \eta \left( \delta_{\alpha\mu} \delta_{\beta\nu}  + \delta_{\alpha\nu} \delta_{\beta\mu} \right)
\nonumber
\\
&=& \lim_{q\to 0}\frac{\mathrm{Im}\chi^{\alpha\beta,\mu\nu}_{\Pi}(\bq,\omega)}{\omega}.
\label{KuboPiFull}
\eea

A very useful formula for the bulk viscosity follows from Eq.~(\ref{KuboPiFull})
by looking at its $(xx,yy)$ component and combining it with the $(xx,xx)$ component in Eq.~(\ref{KuboPiL}).
Using the summation convention, we thus obtain
\bea 
\mathrm{Re}\; \zeta(\omega) = \lim_{q\to 0}\frac{\mathrm{Im}\chi^{\alpha\alpha,\beta\beta}_{\Pi}(\bq,\omega)}{9\;\omega}.
\label{KuboPiZeta}
\eea

We emphasize again that the Kubo formulas for the bulk and shear viscosities expressed in terms 
of the stress-stress correlation function are equivalent to
those expressed in terms of current-current correlations, Eqs.~(\ref{KuboTR}) and (\ref{KuboLR}). 
The two sets of equations are simply related by the exact conservation law, Eq.~(\ref{euler-op}).
Above, we focused on the \emph{dissipative} parts of the response, i.e., the \emph{real} parts of the viscosities.
Comparing Eqs.~(\ref{KuboT}) and (\ref{KuboL}) with Eqs.~(\ref{TPi}) and (\ref{LPi}), we
see that the imaginary part of $\eta$ and the imaginary part of $\left(4\eta/3 + \zeta\right)$ are \textit{not} 
given by $\lim_{\omega\to 0}\lim_{q\to 0}\mathrm{Re}\chi^{xy,xy}_{\Pi}/\omega$ and $\lim_{\omega\to 0}\lim_{q\to 0}\mathrm{Re}\chi^{xx,xx}_{\Pi}/\omega$, respectively.  $\mathrm{Im}\; \eta$ and $\mathrm{Im}\; \zeta$,  
when written in terms of stress correlators, also involve the frequency-independent, equal-time commutator terms in Eqs.~(\ref{TPi}) and (\ref{LPi}). 
This point seems to be missed in treatments that start out with the stress correlator formalism. 
The imaginary parts of the transport coefficients 
are most simply expressed in terms of the current correlation functions, Eqs.~(\ref{KuboT}) and (\ref{KuboL}). 
In the $\omega \to 0$ limit, the validity of this assertion can be seen quite independently from
hydrodynamics (see Appendix~\ref{hydrosec}). Allowing $\eta$ to be complex in the hydrodynamic expression for the transverse current correlation function in Eq.~(\ref{chiT}), for instance, one can readily confirm that the imaginary part of the shear viscosity is indeed given by Eq.~(\ref{KuboT}).  


\section{Shear viscosity and normal fluid density}
\label{rhon-section}

In this Section  we discuss the relation between the static shear viscosity $\mathrm{Re}\; \eta(\omega=0)$
and the normal fluid density $\rho_n$, both of which can be written in terms of the \emph{transverse} current-current
correlation function. This allows us to prove that a non-zero
normal fluid density $\rho_n$ is a necessary condition for a non-vanishing shear viscosity $\eta$.
This is, perhaps, not entirely unexpected on physical grounds, but we are unaware of 
a microscopic proof, valid for all Galilean invariant Bose or Fermi quantum fluids, 
that does not rely on a quasiparticle approximation.

Before turning to the calculation, it may be useful to review some elementary facts
about the shear viscosity $\eta$. Given that there is a Kubo formula for $\eta(\omega)$ 
in terms of the current-current correlation function, Eq.~(\ref{KuboTR}), and that in kinetic theory 
$\eta$ is proportional to the mean free path, it may seem natural to assume
that the shear viscosity of a fluid is the analog of metallic conductivity. 
This, however, is completely misleading. The shear viscosity is, in fact, the analog of the \emph{resistivity}. 
This is clear, e.g., from the classical formula of Poiseuille
for the flow rate $Q = {\pi R^4 \Delta P}/(8 \eta L)$, with a pressure difference $\Delta P$ across
a cylindrical pipe of radius $R$ and length $L$.  
We will see below that zero viscosity in a superfluid is the analog of zero resistance in
a superconductor.

We begin by rewriting the Kubo formula for the shear viscosity, given by Eq.~(\ref{KuboTR}), using the spectral representation in Eq.~(\ref{ImchiAB}):
\bea
\mathrm{Re}\eta(\omega) = \lim_{(T)} \frac{\pi m^2}{\cal{Z}}
\sum_{a,b}\left[e^{-\beta E_a}\;-\;e^{-\beta E_b}\right] E_{ba}
\nonumber\\
\times\frac{\vert \langle b|\hat{j}^x_{\bq}|a\rangle \vert^2}{q^2} 
\delta(\omega -E_{ba}),  
\label{etaomega}
\eea
Here and below, the ``transverse limit'', denoted by $ \lim_{(T)}$, means that for $\chi_J^{xx}$
we first set $q_x = 0$ and then take the limit $q_y \to 0$.

The normal fluid density $\rho_n$ characterizes the response of a fluid to
moving walls and determines the moment of inertia of a cylinder containing the fluid; see, e.g.,
the detailed discussion in Refs.~\cite{Baymbook,NozieresPines2}. It is defined in terms of the
real part of the static transverse current correlator:
\bea
\rho_n = \lim_{q \to 0} m^2 \mathrm{Re}\chi_T(\bq,\omega=0).
\label{rhon1}
\eea
Using the spectral representation in Eq.~(\ref{FourierchiAB}) for $\chi_J^{xx}$, we
can rewrite this result as
\bea
\rho_n = \lim_{(T)} \frac{m^2}{\cal{Z}}
\sum_{a,b}\frac{\left[e^{-\beta E_a}\;-\;e^{-\beta E_b}\right]}{E_{ba}}
{\vert \langle b|\hat{j}^x_{\bq}|a\rangle \vert^2}.
\label{rhon2}
\eea

Our goal now is to understand the connection between the shear viscosity
$\eta$, which is obtained by taking the $\lim_{\omega \to 0}\lim_{q \to 0}$ of $\mathrm{Im}\chi_T$
in Eq.~(\ref{etaomega}), and the normal fluid density $\rho_n$, which is  
the $\lim_{q \to 0}\lim_{\omega \to 0}$ of $\mathrm{Re}\chi_T$ in Eq.~(\ref{etaomega}).
In lattice models of superconductors, it has been suggested \cite{Scalapino} that the order of the
$q$ and $\omega$ limits can be safely interchanged for the \emph{transverse} current correlator,
because all ``transverse'' excitations are gapped (unlike longitudinal excitations such as
phonons in charge-neutral systems). However, this argument is \emph{not} valid for the systems
of interest to us. This can be seen, e.g., from the hydrodynamic form of $\chi_T$ in Eq.~(\ref{chiT}) 
which has a ``diffusion pole'' that makes the order of limits quite different. 

To prove the result stated at the beginning of this Section, we will show that
$\rho_n = 0$ implies $\eta = 0$. The starting condition $\rho_n = 0$ makes sense
only at $T=0$, since at any finite temperature there will necessarily be some thermal excitations.
Furthermore, the vanishing of the normal fluid density 
\bea
\rho_n = \lim_{(T)} 2 m^2
\sum_{b}\frac{\vert \langle b|\hat{j}^x_{\bq}|0\rangle \vert^2}{E_{b0}}
\label{rhon3}
\eea
at $T=0$ implies that each term in the sum $\sum_b$ over states vanishes. 
This means that, for each state $|b\rangle $, if the excitation energy varies as $\lim_{(T)}E_{b0} \sim q^{\alpha_b}$, with
$\alpha_b \geq 0$, then the matrix element of the current operator vanishes even faster:
$\lim_{(T)}\vert \langle b|\hat{j}^x_{\bq}|0\rangle \vert \sim q^{\alpha_b + \beta_b}$
with $\beta_b > 0$. Note that we are not making any assumptions about the nature of the 
spectrum since both gapless ($\alpha_b > 0$) and gapped ($\alpha_b = 0$) excitations are permitted.
In either case, the matrix element of $\hat{j}^x_{\bq}$ vanishes, since the $q \to 0$ limit
of $\hat{j}^x_{\bq}$ is the total momentum, which commutes with the Hamiltonian in a Galilean invariant system.
It is only in such a system that $\rho_n$ vanishes at $T=0$~\cite{NozieresPines2,Paramekanti}.

Now that we have constrained the matrix elements for any form of the excitation spectrum given $\rho_n = 0$,
we now ask how these constraints impact the shear viscosity. We look separately at the contribution from
gapless and gapped states to Eq.~(\ref{etaomega}), which at $T=0$ can be written as
\bea
\eta(\omega) = \lim_{(T)} {\pi m^2}
\sum_{b} E_{b0}\frac{\vert \langle b|\hat{j}^x_{\bq}|0\rangle \vert^2}{q^2} 
\delta(\omega -E_{b0}).
\label{etaomega2}
\eea
Each gapless state $b$, with $\alpha_b > 0$, will contribute a term
$\lim_{(T)} q^{2\alpha_b + \beta_b - 2}\delta(\omega - A_b q^{\alpha_b})$,
which gives a vanishing contribution \cite{eta-zero} in the limit $q \to 0$ for all $\omega > 0$.
Finally taking the $\omega \to 0$ limit, we find that the contribution of the gapless states
to $\eta$ vanishes. 

Next, consider the gapped states with $\alpha_b = 0$, so that 
$\lim_{(T)}E_{b0} \equiv \Delta_b > 0$. Their contribution to Eq.~(\ref{etaomega2})
yields an expression of the form 
$\eta(\omega) = \lim_{(T)} 
\sum_{b}^\prime C_b q^{\beta_b - 2} \delta(\omega -\Delta_{b})$,
where the prime indicates a sum over all gapped states. This result contributes
to both the $\eta$ sum rule and the high-frequency tail that we will
derive later in the paper.  The important point here is that 
for $0< \omega < \min_b^\prime\left\{\Delta_b\right\}$, i.e., below the minimum gap
of all excitations, $\eta(\omega) = 0$. 

Thus, we conclude that 
the vanishing of the normal fluid density implies that the static limit of the shear viscosity
vanishes as well: $\eta = 0$. This means that the Galilean invariant ground state of a superfluid
has zero shear viscosity~\cite{zeroviscosity}. This is similar to the zero d.c. resistivity of a charged superconductor,
as already mentioned at beginning of this Section.
There is, however, an important difference in that the vanishing resistivity persists all the way
up to the transition temperature $T_c$. Even though there are normal fluid excitations in a 
superconductor, the infinite conductivity of the condensate ``shorts out'' the normal fluid in
a superconductor. In marked contrast, in a neutral superfluid, 
$\eta$ vanishes only at $T=0$. For $0 < T < T_c$, even though a condensate exists,
the normal fluid excitations give rise to a non-zero shear viscosity.


\section{Positivity of spectral functions}
\label{posdefsec}

We simplify notation and write from now on
\bea
\eta(\omega) \equiv \mathrm{Re}\; \eta(\omega)
\; \; \mathrm{and} \; \; 
\zeta(\omega) \equiv \mathrm{Re}\; \zeta(\omega),
\label{simplifyReal}
\eea
unless explicitly stated otherwise. This should cause no confusion since we will not be dealing 
with the corresponding imaginary parts.   
Before deriving sum rules for
$\eta(\omega)$ and $\zeta(\omega)$ in Section~\ref{sumrulesec}, 
it is important to discuss here their positivity properties.
Every time we say `positive' we actually mean 
`non-negative', a term we find awkward for repeated use. 
We will show that
\bea 
\eta(\omega)\ge 0\;\; \mathrm{and}\;\; \zeta(\omega) \ge 0
\;\;\;\; \forall \omega.
\label{Work}
\eea
The simplest approach is to make explicit use of the spectral representation. We will see that this
is sufficient to prove the positivity of $\eta(\omega)$, but \emph{not} that of $\zeta(\omega)$.
To prove the latter, we will calculate the power absorbed by the
fluid from an external velocity perturbation with $\nabla \cdot \bu \neq 0$. 

Let us begin with Eqs.~(\ref{KuboTR}) and (\ref{KuboLR}) and
use the spectral representation given by Eq.~(\ref{ImchiAB}) with $\hat{A}_{\bq}=\hat{j}^x_{\bq}$
and $\hat{B}^{\dagger}_{\bq} = \hat{j}^x_{-\bq}$. The transverse and longitudinal components
are obtained, as usual, by taking suitable $q \to 0$ limits~\cite{limits}.
Using $|\langle n|\hat{j}^x_{-\bq}|m\rangle|^2 \geq 0$ and $\omega[1-\exp(-\beta\omega)]\geq 0$ for all $\omega$,
we see that both
$\omega\mathrm{Im}\chi_T(\bq,\omega)$ and $\omega\mathrm{Im}\chi_L(\bq,\omega)$ are positive.
Thus we obtain  
\bea 
\eta(\omega)\ge 0\;\; \mathrm{and}\;\; \zeta(\omega) +4\eta(\omega)/3\ge 0
\;\;\;\; \forall \omega.
\label{Work2}
\eea

The inequality for $\zeta(\omega)$ is much weaker than what we wish to prove.
One reason to expect that a stronger result should exist for $\zeta(\omega)$ is that
it is known from hydrodynamics (see Sec.~49 of Landau and Lifshitz~\cite{LLFM}) that 
the \emph{static} bulk viscosity $\zeta(0)$ must be positive.
To generalize this to all frequencies, we exploit the idea that
the time-averaged power absorbed by the system from an external perturbation
is necessarily positive. 

The rate at which the external velocity perturbation given by Eq.~(\ref{Hpert}) does work on the fluid is given 
by
\bea 
\frac{d W}{dt} = i\omega m\int d\br e^{-i\omega t} e^{0^+t}\bu(\br)\cdot\langle \hat{\bj}(\br,t)\rangle. \label{Power}
\eea
Following Ref.~\cite{ChaikinLubensky}, one finds that the time average of the
power absorbed by the fluid is 
\bea \overline{\frac{dW}{dt}} = \frac{m^2}{2} \sum_\bq 
u_{\alpha}(-\bq)\left[\omega\mathrm{Im}\chi^{\alpha\beta}_J(\bq,\omega)\right]u_{\beta}(\bq)>0.
\label{Power2}\eea
$\overline{dW}/dt > 0$ follows from the fact that energy can only be dissipated 
for \emph{any} choice of the external velocity field.
This implies that the real, symmetric matrix $\omega\mathrm{Im}\chi^{\alpha\beta}(\bq,\omega)$ must be positive definite,
which is equivalent to the positivity of its eigenvalues. 
Using  Eq.~(\ref{chiTL}), we see that these eigenvalues are precisely
$\omega \mathrm{Im}\chi_L(\bq,\omega)$ and $\omega \mathrm{Im}\chi_T(\bq,\omega)$,
so that we simply rederive Eq.~(\ref{Work2}), and do not obtain $\zeta(\omega) \ge 0$.

To constrain $\zeta(\omega)$, without any $\eta(\omega)$ contribution, we must
look at an external velocity field $\bu(\br,t) = \bu(\br)e^{-i\omega t}$ with
$\bu(\br) = a\br$, where $a = \left(\nabla \cdot \bu\right)/3$ is
spatially uniform. 
To analyze the effect of such a perturbation, we first need to rewrite Eq.~(\ref{Power2})
in terms of the stress correlator so that $\partial_\alpha u_\beta$ is directly involved. Second,
$\bu(\br) = a\br$ is not Fourier transformable, so we must work in $\br$-space, rather than
$\bq$-space used elsewhere in the paper.

We use the same derivation that led from the operator Euler equation given by Eq.~(\ref{euler-op}) to Eq.~(\ref{Correlationrelation}), 
to get
\bea 
m^2\omega^2\mathrm{Im}\chi^{\alpha\beta}_{J}(\bq,\omega)  = q_{\mu}q_{\nu}\mathrm{Im}\chi^{\alpha\mu,\beta\nu}_{\Pi}(\bq,\omega). 
\label{ImCorrelationrelation}
\eea
Using this in Eq.~(\ref{Power2}) and rewriting the resulting expression in real space,  we get
\bea 
\lefteqn{\overline{\frac{dW}{dt}} = \frac{1}{2}\int \!d\br \!\int \!d\br' \times}&&\nonumber\\&&\partial_{\alpha}u_{\mu}(\br)\left[\frac{\mathrm{Im}\chi^{\alpha\mu,\beta\nu}_{\Pi}(\br-\br',\omega)}{\omega}\right]\!\partial_{\beta}u_{\nu}(\br'), 
\label{Power3} 
\eea
which must hold for arbitrary velocity fields $\bu(\br)$.

To isolate the contribution of the bulk viscosity, we choose the velocity field $\bu = a\br$, 
for which the shear term (in square brackets) 
in the viscous stress tensor, Eq.~(\ref{stresstensor}), vanishes.
Using $\partial_\alpha u_\beta = a \delta_{\alpha\beta}$ in Eq.~(\ref{Power3}) we get 
$\mathrm{Im}\chi^{\alpha\alpha,\beta\beta}_{\Pi}(\bq \to 0,\omega)/\omega \ge 0$. From the result
given by Eq.~(\ref{KuboPiZeta}) for the bulk viscosity, it immediately follows that $\zeta(\omega)\ge 0$ for all $\omega$.  

\section{Sum rules}
\label{sumrulesec}
 
We now derive sum rules for the shear and bulk viscosities, 
$\int^{\infty}_0 d\omega \eta(\omega)$ and $\int^{\infty}_0 d\omega \zeta(\omega)$.
We will first show that 
\bea 
\lefteqn{\frac{1}{\pi}\int^{\infty}_0 \!\!d\omega
\lim_{q\to 0}\!\frac{\omega}{q^2}\mathrm{Im}\chi^{xx}_{J}(\bq,\omega)=}&&\nonumber
\\&&\!\!\!\!\!\!\!\!\lim_{q\to 0}
\frac{\langle[\hat{j}^{x}_{-\bq},[\hat{H},\hat{j}^{x}_{\bq}]]\rangle}{2 q^2}
+\lim_{\omega\to 0}\lim_{q\to 0}\frac{\omega^2}{2q^2}\mathrm{Re}\chi^{xx}_{J}(\bq,\omega).
\label{noncommutator}
\eea
Then we will simplify the two terms on the right hand side of Eq.~(\ref{noncommutator}):
the first term by explicit evaluation of the commutators, and the second by appealing to
hydrodynamics.
   
To see what is involved in deriving Eq.~(\ref{noncommutator}), let us first be na\"ive and
ignore the $q \to 0$ limit.
Evaluating the integral on the left hand side by using the
spectral representation in Eq.~(\ref{ImchiAB}) for $\mathrm{Im}\chi^{xx}_{J}$,
we only obtain the first commutator term on the right.
But taking the $q \to 0$ limit after doing the $\omega$-integration
is \emph{not} the same as interchanging the order of these operations!
In order to do it correctly ($q \to 0$ limit before the $\omega$-integration),
we exploit the Kramers-Kronig (K-K) relations to evaluate the 
integral in Eq.~(\ref{noncommutator}).
The only subtle point in this approach is that we need to ensure that the 
analytic functions which we K-K transform
decay sufficiently rapidly for $\omega \to \infty$.

Using the expression in Eq.~(\ref{FourierchiAB}), it is straightforward to expand 
the current correlator in powers of $\omega^{-1}$ for large frequencies.  
One finds~\cite{Pitaevskiibook},
\bea 
\lim_{\omega\to \infty}\chi^{xx}_J(\bq,\omega) &=&  \frac{\langle[\hat{j}^{x}_{-\bq},\hat{j}^{x}_{\bq}]\rangle}{\omega} - \frac{\langle[\hat{j}^{x}_{-\bq},[\hat{H},\hat{j}^{x}_{\bq}]]\rangle}{\omega^2} + \ldots.
\nonumber\\ 
\label{chiJasymptote}
\eea
The $\omega^{-1}$ term vanishes since $\langle[\hat{j}^{x}_{-\bq},\hat{j}^{x}_{\bq}]\rangle = -(2q_{x}/m^2)\sum_{\bk\sigma}n_{\bk\sigma} k_{x} = 0$ in a uniform system.  
We further note that this expansion is strictly valid only for a smooth potential~\cite{divergence}, a point
which we will elaborate on in later Sections. 
  
Let us define a function $F(\omega)$ as
\bea F(\omega) \equiv \lim_{q\to 0}\frac{\omega^2}{q^2}\left[\chi^{xx}_J(\bq,\omega) +  \frac{\langle[\hat{j}^{x}_{-\bq},[\hat{H},\hat{j}^{x}_{\bq}]]\rangle}{\omega^2}\right],
\label{F} \eea
where the $q \to 0$ limit is defined appropriately~\cite{limits} for the longitudinal and
transverse cases.
From Eq.~(\ref{chiJasymptote}), we see that $\lim_{\omega\to\infty}F(\omega)$ vanishes
at least as fast as $\omega^{-1}$ and we can K-K transform it. We thus obtain
\bea 
\lim_{\omega\to 0}\mathrm{Re} F(\omega) &=& \frac{{\cal{P}}}{\pi}\int^{\infty}_{-\infty}d\omega'\frac{\mathrm{Im}F(\omega')}{\omega'}\\
&=&\frac{2}{\pi}\int^{\infty}_{0}d\omega'
\lim_{q\to 0}\frac{\omega'}{q^2}\mathrm{Im}\chi^{xx}_J(\bq,\omega').\nonumber
\label{noncommutator2}
\eea
where we have used the fact that $\omega \mathrm{Im}\chi^{xx}_J(\bq,\omega)$ is an even function
of $\omega$.
Using Eq.~(\ref{F}) on the left-hand side of this expression immediately leads to
the result, Eq.~(\ref{noncommutator}), quoted above.  

As mentioned earlier, $\lim_{\omega\to 0}\lim_{q\to 0}(\omega^2/2q^2)\mathrm{Re}\chi^{xx}_J$ 
in Eq.~(\ref{noncommutator}) arises from the noncommutativity of the $\omega\to 0$ and $q\to 0$ limits.  
Since this term involves the zero-frequency, long-wavelength limit where hydrodynamics is applicable, we can use hydrodynamic expressions for the current correlation function to evaluate it.  
In Appendix~\ref{hydrosec}, we review such expressions and show that for any simple hydrodynamic liquid, one has
\bea 
\lim_{\omega\to 0}\lim_{q\to 0}\frac{m^2\omega^2}{2q^2}\mathrm{Re}\chi_T(\bq,\omega) &=& 0,\nonumber\\
\label{difference2}\\
\lim_{\omega\to 0}\lim_{q\to 0}\frac{m^2\omega^2}{2q^2}\mathrm{Re}\chi_L(\bq,\omega) &=& - \frac{\rho c_s^2}{2},\nonumber 
\eea
where the adiabatic sound speed is $c_s \equiv (\partial P/\partial\rho)^{1/2}$ at fixed $s=S/N$. 
Equation~(\ref{difference2}) is valid for both normal fluids and superfluids 
(within two-fluid hydrodynamics).

Combining Eqs.~(\ref{KuboTR}), (\ref{KuboLR}), (\ref{noncommutator}), and (\ref{difference2}), we find the following sum rules:
\bea \frac{1}{\pi}\int^{\infty}_0 \!\!d\omega
\eta(\omega)=\lim_{q\to 0}\frac{m^2\langle[\hat{j}^{x}_{-\bq},[\hat{H},\hat{j}^{x}_{\bq}]]\rangle_T}{2 q^2},\label{etasumrule}\eea
\bea \frac{1}{\pi}\!\int^{\infty}_0 \!\!\!d\omega\!\!
\left[\!\zeta(\omega)\!+\!\frac{4\eta(\omega)}{3}\right] \!=\! \lim_{q\to 0}\frac{m^2\langle[\hat{j}^{x}_{-\bq},[\hat{H},\hat{j}^{x}_{\bq}]]\rangle_L}{2 q^2}\! -\! \frac{\rho c_s^2}{2}. \nonumber\\ \label{zetasumrule}\eea
Here, $\langle\cdots\rangle_{T(L)}$ denotes the $q \to 0$ limit appropriate to the transverse (longitudinal) case~\cite{limits}.   

The last remaining step in our derivation is  
to evaluate the commutators in Eqs.~(\ref{etasumrule}) and (\ref{zetasumrule}).
We consider a system of fermions or bosons described by the Hamiltonian
\bea 
\hat{H}\! &=& \hat{K} + \hat{V}
\\
&=& \! \sum_{\bk\sigma}\varepsilon_{\bk}\hat{c}^{\dagger}_{\bk\sigma}\hat{c}_{\bk\sigma}\! +\!\frac{1}{2}\!\sum_{\substack{\bk\bk'\bp\\ \sigma\sigma'}}\!V(p)\hat{c}^{\dagger}_{\bk+\bp\sigma}\hat{c}^{\dagger}_{\bk'\!-\bp\sigma'}\hat{c}_{\bk'\!\sigma'}\hat{c}_{\bk\sigma}.\nonumber
\label{H}\eea
For a single-component Bose system, $\sigma=\sigma'$ assumes one value; 
for fermions, $\sigma = \uparrow,\downarrow$ can take one of two ``spin'' values.  
It is straightforward, but tedious, to evaluate the commutator in Eq.~(\ref{noncommutator}) for this Hamiltonian.  
One finds, for both fermions and bosons, 
\bea
\lefteqn{\frac{m^2}{2}\langle[\hat{j}^{x}_{-\bq},[\hat{H},\hat{j}^{x}_{\bq}]]\rangle=}&&
\label{Jsumrule}
\\
&&\frac{\langle \hat{K} \rangle}{3}\left(2q^2_{x} + q^2\right) + n\frac{ q^2q^2_{x}}{8m}- 
\frac{1}{2} \langle\!\langle 2V(p)p^2_{x} 
\nonumber
\\
&-&V\!(|\bp-\bq|)(p_{x}\!-\!q_{x})^2 
-  V\!(|\bp+\bq|)(p_{x}\!+\!q_{x})^2 \rangle\!\rangle.  
\nonumber
\eea
Here, $\langle \hat{K} \rangle = \sum_{\bk\sigma}\varepsilon_{\bk} n_{\bk\sigma}$ is the kinetic energy density,
and we have introduced the shorthand notation 
\bea 
\left\langle\!\left\langle {\cal Q} \right\rangle\!\right\rangle \equiv \frac{1}{2}\sum_{\substack{\bk\bk'\bp\\ \sigma\sigma'}}
{\cal Q} \langle \hat{c}^{\dagger}_{\bk+\bp\sigma}\hat{c}^{\dagger}_{\bk'-\bp\sigma'}
\hat{c}_{\bk'\sigma'}\hat{c}_{\bk\sigma}\rangle.
\label{intaverage}
\eea
Related expressions specific to Bose liquids are given in Ref.~\cite{Dalfovo92}.  We also note in passing that the longitudinal component of Eq.~(\ref{Jsumrule}) is related by Eq.~(\ref{Nconserv}) to the so-called ``$\langle \omega^3\rangle $ sum rule" discussed for electronic systems~\cite{Puff65}.  

The right-hand side of Eq.~(\ref{Jsumrule}) varies as $q^2$ as $q\to 0$, 
which cancels the $1/q^2$ in Eqs.~(\ref{etasumrule}) and (\ref{zetasumrule}).
Evaluating the transverse and longitudinal limits of Eq.~(\ref{Jsumrule}), one finds the following viscosity sum rules:
\bea 
\frac{1}{\pi}\int^{\infty}_{0}\!\!d\omega \eta(\omega)= \frac{\varepsilon}{3}-\frac{\langle \hat{V} \rangle}{3} + \frac{2\overline{V}'}{15} 
+ \frac{\overline{V}''}{30}
\label{etasumruleiso}
\eea
and 
\bea 
\frac{1}{\pi}\int^{\infty}_{0}\!\!d\omega \zeta(\omega) = 
\frac{5\varepsilon}{9}+\frac{4\langle \hat{V} \rangle}{9} +
\frac{5\overline{V}'}{9}+ 
\frac{\overline{V}''}{18}-\frac{\rho c_s^2}{2}.
\label{zetasumruleiso}
\eea
Here,
$\varepsilon=\langle \hat{K} \rangle+\langle \hat{V} \rangle$ is the total energy density, 
$\langle \hat{V} \rangle$ the potential energy density, and the terms
$\overline{V}'$ and $\overline{V}''$ are defined using Eq.~(\ref{intaverage}) as
\bea 
\overline{V}' \equiv \left\langle\!\left\langle p\left({\partial V}/{\partial p}\right) \right\rangle\!\right\rangle \ \ 
{\rm and} \ \
\overline{V}'' \equiv \left\langle\!\left\langle p^2\left({\partial^2V}/{\partial p^2}\right) \right\rangle\!\right\rangle.
\label{Vpp}
\eea
  
These sum rules are valid at all temperatures (i.e., in the superfluid as well as normal phase) for any Bose or Fermi system with an arbitrary, 
spin-independent, isotropic interaction potential $V(p)$. We emphasize that these are exact results obtained without making
any quasiparticle approximations.
In the next Section (Sec.~\ref{diluteviscositysec}), we simplify these sum rules 
for the case of a two-component Fermi gas with short range interactions,
which is of relevance to experiments on ultracold atomic Fermi gases with Feshbach scattering resonances.

Before closing this Section, let us briefly discuss viscosity sum rules using the stress correlator representation.
For the shear viscosity spectral function, the sum rule
\bea 
\int^{\infty}_0 d\omega \eta(\omega) &=& \frac{1}{\pi}\int^{\infty}_0 d\omega \lim_{q\to 0} 
\frac{\mathrm{Im}\chi^{xy,xy}_{\Pi}(\bq,\omega)}{\omega} \nonumber\\ &=& 
\lim_{\omega\to 0}\lim_{q\to 0}\frac{1}{2}\mathrm{Re}\chi^{xy,xy}_{\Pi}(\bq,\omega)
\label{etasumrulePi}\eea
follows trivially from the Kramers-Kronig relation.
To show that this is the same as Eq.~(\ref{etasumrule}), we
use Eq.~(\ref{TPi}) in the second line of Eq.~(\ref{etasumrulePi}).  
One can rewrite the commutator in Eq.~(\ref{TPi}) using the Fourier transform of 
Eq.~(\ref{euler-op}), and set $\omega^2 \mathrm{Re}\chi_T/q^2$ to zero
using the hydrodynamic result, Eq.~(\ref{difference2}), to obtain Eq.~(\ref{etasumrule}).

\section{Dilute two-component Fermi gas}
\label{diluteviscositysec}

We now specialize to the case of a two-component Fermi gas in the dilute limit, where the effective range $r_0$ of the 
potential (van der Waals at ``long" distances, with $r_0\sim 100 a_0$) is much smaller than the $s$-wave scattering length $a$ and the mean interparticle spacing $k^{-1}_F$. (In typical experiments, $k^{-1}_F\sim 1\mu m$ and $500 a_0\lesssim |a| \lesssim \infty$.)
In the zero range limit $r_0\to 0$, all physical observables are
universal ($r_0$-independent) functions of the 
energy scale $\epsilon_F$ (or length scale $k^{-1}_F$) and the dimensionless parameters
$T/\epsilon_F$ (temperature) and $1/(k_F a)$ (interaction).   
We will show that for Fermi gases, the results given by
Eqs.~(\ref{etasumruleiso}) and (\ref{zetasumruleiso}) of the previous Section,  
reduce to the simple expressions given by Eqs.~(\ref{etasumrule0finite}) and (\ref{zetasumrule0})
in the Introduction. 
 
Our main task is to calculate the terms $\overline{V}'$ and $\overline{V}''$, involving \textit{gradients} 
of the interaction potential, defined in Eq.~(\ref{Vpp}). 
We use the real-space approach developed by Zhang and Leggett~\cite{Zhang08}, which is a simple
way to derive results first obtained by Tan~\cite{Tan08,Braaten08}. 
Using the two-body density matrix
\bea
\lefteqn{{\cal F}(\br) = }&&\nonumber\\&&\!\!\!\!\!\int\!\! d^3\bR \Big\langle \hat{\psi}^{\dagger}_{\uparrow}(\bR\!+\!\frac{\br}{2})
\hat{\psi}^{\dagger}_{\downarrow}(\bR\!-\!\frac{\br}{2})
\hat{\psi}_{\downarrow}(\bR\!-\!\frac{\br}{2})\hat{\psi}_{\uparrow}(\bR\!+\!\frac{\br}{2})\Big\rangle
\nonumber
\\
\label{rho2def}
\eea
we rewrite $\overline{V}'$ and $\overline{V}''$ in real space as
\bea 
\overline{V}' = \int d^3\br \; r \frac{\partial V(r)}{\partial r} {\cal F}(\br)
\label{moment1} 
\eea 
and
\bea 
\overline{V}''
= \int d^3\br\; r^2 \frac{\partial^2 V(r)}{\partial r^2} {\cal F}(\br).
\label{moment2}
\eea 
Since $V(r)$ is short-ranged, these expressions are only sensitive to the short-distance ($r_0\lesssim r \ll k^{-1}_F$) structure of the two-body density matrix.  (The non-universal contribution from distances smaller than $r_0$ is assumed to be small.)  For a  two-component dilute Fermi gas, at these short distances, the two-body density matrix is~\cite{Zhang08}
\bea 
{\cal F}(\br) = \frac{C}{16\pi^2}\left(\frac{1}{r}-\frac{1}{a}\right)^2.
\label{rho2}
\eea
Here, $C$ is the contact~\cite{Tan08,Zhang08,Braaten08} mentioned in the Introduction. 
In Appendix \ref{ContactAppendix} we remind the reader how the contact $C$ governs both the short-distance
behavior of the two-body density matrix in Eq.~(\ref{rho2}), and the 
large-$k$ tail of the momentum distribution function
$\lim_{k\to \infty} n_{\bk\sigma} = {C}/{k^4}$.

Using integration by parts, we transform gradients of the potential $V(r)$ 
in Eqs.~(\ref{moment1}) and (\ref{moment2}) into gradients of the 
two-body density matrix, Eq.~(\ref{rho2}). We thus find
\bea 
\overline{V}' = \frac{C}{4\pi}\int dr V(r)\left(-1 + {4r}/{a}\right)\label{moment1b}
\eea
and
\bea \overline{V}'' =\frac{C}{2\pi}\int dr V(r)\left(1 - {6r}/{a}\right).\label{moment2b}
\eea
All that remains is to evaluate the two integrals
$X_n =  C\int dr V(r) (r/a)^n/4\pi$ with $n=0,1$ in the limit
where the range of the potential $r_0 \to 0$. The Tan relations
are precisely what we need to evaluate such (possibly divergent) integrals. 
The details of this analysis are described in Appendix \ref{ContactAppendix}.
We use the potential energy density~\cite{Tan08,Braaten08}
\bea 
\langle \hat{V} \rangle  = - \frac{C \Lambda}{2\pi^2 m} + \frac{ C}{4\pi m a},
\label{eint}
\eea
where $\Lambda \equiv 1/r_0$ is the ultraviolet cutoff ,
and the pressure
\bea 
P = 2\varepsilon/3 + {C}/(12\pi m a)
\label{tanP}
\eea
to determine $X_0$ and $X_1$. 
In deriving these results, we also use an expression for the pressure 
$P$ in terms of $\varepsilon$, $\langle \hat{V} \rangle$ and $\overline{V}'$
which is derived in Appendix~\ref{Pressuresec} using the Feynman-Hellmann theorem.

Our final results for $\overline{V}'$ and $\overline{V}''$, 
derived in Appendix \ref{ContactAppendix}, are
\bea 
\overline{V}' = 
-\langle \hat{V} \rangle -2\varepsilon + 3P = C\Lambda/{2\pi^2 m}
\label{identity1}
\eea
and
\bea \overline{V}'' 
= 2\langle \hat{V}\rangle + 8\varepsilon - 12P = - \frac{C \Lambda}{\pi^2 m} - \frac{ C}{2\pi ma} 
\label{identity2}
\eea

Using these results in the general sum rules given by Eqs.~(\ref{etasumrule}) and (\ref{zetasumrule}), we
obtain the $\eta$ and $\zeta$ sum rules for the two-component dilute
Fermi gas which are valid for \emph{all} values of $1/(k_F a)$ throughout the BCS-BEC crossover, 
so long as $a,k^{-1}_F\gg r_0$, and at all temperatures, both in the superfluid and normal phases, so long as $T\ll 1/mr^2_0$. 
For the shear viscosity, we find
\bea
\int^{\Lambda^2/m}_{0}d\omega\; \eta(\omega)/{\pi} &=& {\varepsilon}/{3} - {2\langle \hat{V} \rangle}/{5}
\nonumber
\\
&=& \frac{\varepsilon}{3} - \frac{C}{10 \pi m a} + \frac{C \Lambda}{5\pi^2 m},
\label{etasumrule2}
\eea
where we have imposed the energy cutoff $\Lambda^2/m = 1/mr_0^2$~\cite{uv-cutoff}.
In the zero-range limit as $\Lambda = 1/r_0 \to \infty$, the right hand side diverges.
(Strictly speaking, every physical potential has a small non-zero effective range $r_0$,
which leads to a well-defined, finite results, but one that is ``non-universal'' in that it depends
on short distance physics.) We will see in the following Section, Sec.~\ref{highfrequencytailsec}, 
how to make sense of this divergence and find a modified sum rule that remains finite as $r_0\to 0$.

For the bulk viscosity we find
\bea 
\int^{\infty}_{0}d\omega\; \zeta(\omega)/{\pi} &=& P - \varepsilon/9 - {\rho c_s^2}/{2}
\\
&=& \frac{5\varepsilon}{9} + \frac{C}{12 \pi m a} - \frac{\rho c_s^2}{2}.
\label{zetasumrule2}
\eea
Below the superfluid transition, the bulk viscosity $\zeta$ that enters Eq.~(\ref{zetasumrule0}) is the bulk viscosity $\zeta_2$,
as explained earlier.
We can rewrite the right hand side of this sum rule in a useful way using
simple facts about the scaling form of thermodynamic functions across the entire BCS-BEC crossover,
as described in detail in Appendix~\ref{universalthermosec}. The final result is
\bea 
\int^{\infty}_{0}d\omega\; \zeta(\omega)/{\pi} = 
\frac{1}{72 \pi m a^2} \left( \frac{\partial C}{\partial a^{-1}} \right)_s,
\label{zetasumrule2A}
\eea
where the derivative is taken at constant entropy per particle $s = S/N$.
The positivity of the sum rule, given that of its integrand, implies that the contact is a 
monotonically increasing function of $1/(k_Fa)$ through the BCS-BEC crossover.  
We will discuss this further in Section~\ref{Crossoversec}.


\section{High-frequency tails}
\label{highfrequencytailsec}

In this Section we derive a  modified shear viscosity sum rule that is manifestly finite
in the $\Lambda = 1/r_0 \to \infty$ limit. This is obtained by relating the 
linear (in $\Lambda$) divergence in the sum rule, Eq.~(\ref{etasumrule2}), to a high-frequency tail 
in $\eta(\omega)\sim 1/\sqrt{\omega}$, and then ``subtracting out'' the contribution of this tail. 
We use ``high frequency'' or $\omega \to \infty$, to mean
$\epsilon_F \ll \omega \lesssim 1/mr_0^2$. 
We also argue that a high-frequency tail of the form
$\omega^{-n/2}$, with odd integer $n$, in a variety of spectral functions
is a generic feature of short-range physics. As discussed below, it shows up in many contexts,
even outside dilute quantum gases.  

We can rewrite the $\eta$ sum rule in Eq.~(\ref{etasumrule2}) as
\bea 
\frac{1}{\pi}\int^{\infty}_{0}\!\!d\omega \left[\eta(\omega) -\frac{C\Theta(\omega - \Omega_0)}{10\pi\sqrt{m\omega}}\right]
\nonumber \\
 = \frac{\varepsilon}{3} - \frac{ C}{10\pi m a} + \frac{C}{5\pi^2}\sqrt{\frac{\Omega_0}{m}},
\label{etasumrule3}
\eea
where $\Omega_0$ is an arbitrary energy scale. If we choose $\Omega_0$ to be $\Lambda^2/m$
we recover Eq.~(\ref{etasumrule2}). But for any finite $\Omega_0$,
subtracting out the $\omega^{-1/2}$ tail makes the integral ultraviolet
convergent and we can take the cutoff $\Lambda$ to infinity.
If we choose $\Omega_0 = 0$, we obtain the
finite sum rule 
\bea 
\frac{1}{\pi}\int^{\infty}_{0}\!\!d\omega \left[\eta(\omega) -\frac{C}{10\pi\sqrt{m\omega}}\right]
 = \frac{\varepsilon}{3} - \frac{ C}{10\pi m a}.  
\label{etasumrule3A}
\eea
The price we pay for using this finite, $r_0$-independent result
(in the $r_0 \to 0$ limit) is that we sacrifice the positivity
of the integrand. At sufficiently small $\omega$, we must necessarily have 
$\eta(\omega) < C/(10\pi\sqrt{m\omega})$ since $\eta(0)$ is finite.
One can, in principle, exploit the freedom in Eq.~(\ref{etasumrule3}) and 
choose $\Omega_0$ to be large enough so 
that the integrand is always positive, however. 

The finiteness of the right hand side of Eq.~(\ref{etasumrule3A}) 
implies that the integrand on the left must vanish at least as fast as $\omega^{-3/2}$ for the integral
to converge at large $\omega$. 
Thus the asymptotic behavior of the spectral function
$\eta(\omega)$ is of the form
\bea 
\eta(\omega \to \infty) \simeq \frac{C}{10\pi\sqrt{m\omega}}.
\label{etahighomega}
\eea

We note that a high-frequency tail in the imaginary part of a 
retarded correlation function which goes like
$\omega^{-n/2}$, with positive integer $n$, is a general feature of short-range two-body physics.  
Suppose that for some operator $\hat{A}$, the corresponding $n$-th moment sum rule has the form 
\bea 
\frac{1}{\pi}\int^{\infty}_0 d\omega \omega^n \mathrm{Im}\chi_{A,A}(\omega) = \alpha \langle V\rangle + \cdots,\label{sumrulediv}
\eea 
where we only show the divergent term explicitly;
the ellipses denote regular terms.  $\alpha$ is some combination of parameters and is not, in general, dimensionless.  
In addition to the current correlation function ($n=1$), diverging sum rules of the form given by Eq.~(\ref{sumrulediv}) arise for the radio frequency (RF) spectral function ($n=1$)~\cite{RFsumrule}, and, as we show below, the density response function ($n=3$).   Using the same reasoning as above, a divergence of the form given by Eq.~(\ref{sumrulediv})
implies a high-frequency tail.  
For a dilute two-component Fermi gas with $a\gg r_0$,
the high-frequency tail is given by
\bea 
\mathrm{Im}\chi_{A,A}(\omega \to \infty) \simeq \frac{\alpha C}{4\pi m^{1/2}}\frac{1}{\omega^{n+1/2}}. 
\eea 

As seen from the above arguments, an $\omega^{-3/2}$ tail arises in the
radio-frequency spectroscopy response function $I(\omega)$~\cite{Schneider09,Strinati09} for Fermi gases. 
Another interesting example is the $\omega^{-7/2}$ tail in the density response of a dilute Fermi gas which we derive
in Section~\ref{SqomegaSec}.
There, we also point out that an identical asymptotic behavior is found for the dense Bose liquid $^4$He, which further
emphasizes the generality of the short-distance physics in all quantum fluids.

\section{Sum rules through the BCS-BEC crossover}
\label{Crossoversec}

In this Section we consider the bulk and shear viscosity sum rules through the BCS-BEC crossover, going from the weakly attractive BCS limit
($a$ small and negative) with large Cooper pairs to the BEC limit ($a$ small and positive) with weakly interacting, tightly bound molecules.  
The crossover can be traversed by changing $x = 1/(k_F a)$ from $x = -\infty$ (BCS limit) to $x = +\infty$ (BEC limit).
In experiments, the scattering length $a$ is varied by tuning a magnetic field about a Feshbach resonance.  Precisely at resonance, $x=0$, the
scattering length diverges and the Fermi gas is in a very strongly interacting ``unitary regime" where the pair size is of the order of the interparticle
spacing. 

To actually compute the viscosity sum rules given by Eqs.~(\ref{zetasumrule2A}) and (\ref{etasumrule3A}) 
for arbitrary coupling $x = 1/(k_F a)$ and temperature $T$,
we need to know the energy density $\varepsilon = n \epsilon_F {\cal{E}}(x,T/\epsilon_F)$, from which we can determine the contact $C$ as described below.
In general, we will need to use quantum Monte Carlo (QMC) data for the energy density to evaluate the sum rules.
However, as shown below, we are able to analytically constrain the bulk viscosity spectral function at unitarity.

\begin{figure}
\begin{center}
\epsfig{file=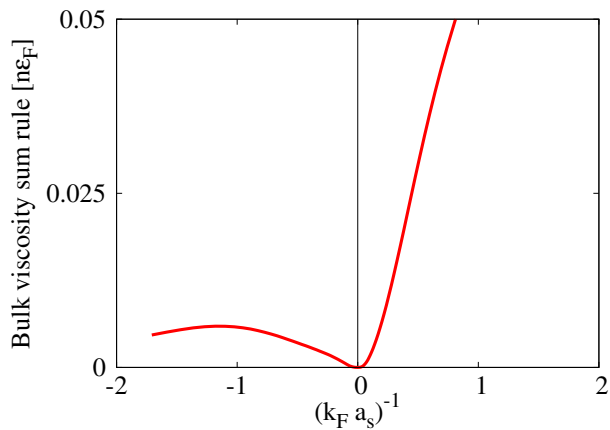, angle=0,width=0.47\textwidth}
\caption{(Color
  online) The  value of the bulk viscosity sum rule given by the right-hand side of Eq.~(\ref{zetasumrule2A}) at $T=0$ in units of $n\epsilon_F$ through the BCS-BEC crossover. }
\label{bulksumrulefig}
\end{center}
\end{figure}

We see from Eq.~(\ref{zetasumrule2A}) that the bulk viscosity sum rule vanishes at unitarity:
\bea 
\frac{1}{\pi}\int^{\infty}_{0}d\omega\; \zeta(\omega)= 0 \ \ \ \ (|a| = \infty).
\label{ZetaSumUnitarity}
\eea
We are using here the fact that $(\partial C / \partial a^{-1})_s$ is finite (i.e., non-infinite) at
$x = 1/(k_F a) = 0$ at all temperatures. One can also see this using
elementary arguments that do not involve the contact. From ``universal thermodynamics"~\cite{Ho04}, the only energy scales at unitarity are $\epsilon_F$ and the temperature, and 
we can directly show that $P - \varepsilon/9- \rho c_s^2/2=0$ (see Appendix~\ref{universalthermosec}).

The vanishing sum rule, Eq.~(\ref{ZetaSumUnitarity}), together with the positivity condition
$\zeta(\omega) \geq 0$ derived in Section~\ref{posdefsec}, implies
\bea
\zeta(\omega) = 0 \ \ \ \forall \omega \ \ \ (|a| = \infty).
\label{ZetaUnitarity}
\eea 
That the \emph{static} bulk viscosity $\zeta(0)$ vanishes is a well-known
consequence~\cite{Son07} of scale or conformal invariance at unitarity~\cite{Castin04}. Our result generalizes this to arbitrary frequencies.  
As discussed below in Section~\ref{SqomegaSec}, our result actually has important implications for measuring the 
frequency dependent shear viscosity of a unitary Fermi gas using a density probe such as two-photon Bragg scattering. 

Another general consequence of $\zeta(\omega) \geq 0$ is that its sum rule must be positive for
all $x = 1/(k_F a)$ and $T$.  Equation~(\ref{zetasumrule2A}) then implies that
\bea 
\left(\frac{\partial C}{\partial a^{-1}}\right)_{{s}}\geq 0\;\;\;\;\forall a,
\eea
so that the contact must be a monotonically increasing function of $1/(k_Fa)$ through the BCS-BEC crossover 
at fixed entropy per particle. We can understand this inequality intuitively as follows: the contact $C$, which
is related to the probability of finding two particles of opposite spin close to each other, can only increase with
increasing attraction $a^{-1}$.
  
In Fig.~\ref{bulksumrulefig} we show the bulk viscosity sum rule in Eq.~(\ref{zetasumrule2A}) at $T=0$ 
calculated using QMC data~\cite{Astrakharchik04} for the energy density $\varepsilon$.
The contact $C$ is obtained from $\varepsilon$ using Tan's ``adiabatic relation"~\cite{Tan08} 
\bea
\left(\partial \varepsilon / \partial a^{-1} \right)_{s} =  - C/(4\pi m),
\label{tanA}
\eea 
where the derivative is taken at fixed entropy per particle $s\equiv S/N$.
We fitted the QMC data and took numerical derivatives with respect to $a^{-1}$.
Since the $\zeta$ sum rule involves the second derivative of QMC data
for the energy density, the results may not be very accurate far from unitarity in either direction.

\begin{figure}
\begin{center}
\epsfig{file=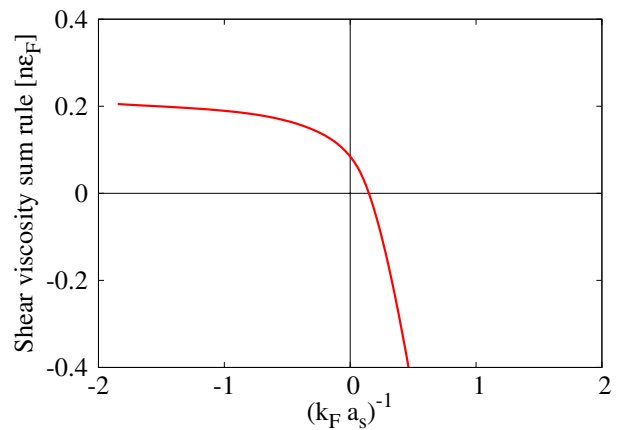, angle=0,width=0.47\textwidth}
\caption{(Color online) The  value of the finite shear viscosity sum rule, with the contribution from the high-frequency tail in Eq.~(\ref{etahighomega}) subtracted out, given by the right-hand side of Eq.~(\ref{etasumrule3A}) at $T=0$ in units of $n\epsilon_F$ through the BCS-BEC crossover.}
\label{shearsumrulefig}
\end{center}
\end{figure}

Both the vanishing of the $\zeta$ sum rule at $x= 1/(k_F a) =0$ and its positivity away from
unitarity are apparent in Fig.~\ref{bulksumrulefig}. This is due to the 
$1/a^2$ dependence of the sum rule in the vicinity of unitarity.
We emphasize the nontriviality of the result given by Eq.~(\ref{zetasumrule2A}) in the unitarity region.
In the form first derived in Eq.~(\ref{zetasumrule2}), the right hand side is
$(P - \varepsilon/9- \rho c_s^2/2)$. Each term in this expression has both constant and order $x$
contributions, which must all cancel to give a final result which goes like $x^2$ at small $x$.
In the BCS limit, the $\zeta$ sum rule vanishes as $2n\epsilon_F/(27\pi|x|$) since $C\to 4\pi^2n^2a^2_s$~\cite{Tan08}.  In the BEC limit, the energy density is dominated by the negative molecular binding energy, $\varepsilon\approx nE_b/2$,  with $E_b = -1/(ma^2)$.  Thus, $C\to 4\pi n/a$ and the sum rule grows as $n|E_b|/18$.

Next, in Fig.~\ref{shearsumrulefig}, we plot the shear viscosity sum rule given by Eq.~(\ref{etasumrule3A}) at $T = 0$ 
again using the QMC data of Ref.~\cite{Astrakharchik04}. 
Because of the $1/\sqrt{\omega}$ subtraction extending all the way down to $\omega = 0$, the $\eta$ sum rule
in Eq.~(\ref{etasumrule3A}) is \emph{not} constrained to be positive.  Using the above analytic result for the contact in the BCS limit, one finds that the $\eta$ sum rule asymptotes to $0.2 n\epsilon_F$
in the BCS limit. 
At unitarity, $|a| = \infty$ and the $\eta$ sum rule is $\varepsilon/3 \simeq 0.4 \times (3 n\epsilon_F/5) \times (1/3) = 0.08 n\epsilon_F$. 
On the BEC side of the resonance the sum rule changes sign, tending to $(17/30)nE_b$ in the BEC limit.


\section{Dynamic Structure Factor}
\label{SqomegaSec}

We now discuss the connection between viscosity and the density-density correlator or dynamic structure factor.
This analysis leads to two interesting results for the two-component Fermi gas. First, we predict that 
a density probe such as two-photon Bragg spectroscopy~\cite{Iacuppo} can in principle be used to measure the
frequency dependent $\eta(\omega)$ at unitarity:
\bea 
\eta(\omega)
= \lim_{q\to 0}\frac{3\omega^3}{4 q^4}\;\mathrm{Im}\chi_{\rho\rho}(\bq,\omega) \ \ \ \ \ (|a| = \infty).
\label{bragg}
\eea
Second, we derive the high-frequency tail~\cite{Son10} 
\bea 
\lim_{\omega\to\infty}\lim_{q\to 0}S(\bq,\omega) = \frac{2q^4C}{15 \pi^2 m^{1/2}}\frac{1}{\omega^{7/2}},
\label{sqomegatail}
\eea
a result that is valid for all $1/(k_F a)$ and all temperatures. As discussed below, such non-analytic
tails are also known in other strongly interacting quantum fluids like $^4$He.

We start with the operator form of the continuity equation 
\bea 
i [\hat{\rho},\hat{H}] = m \partial_{\alpha}\hat{j}_{\alpha},
\label{continuity-op}
\eea
where $\hat{\rho} = m\hat{n}$ is the mass density operator, and take its matrix elements 
between exact many-body eigenstates. Using the spectral representation, Eq.~(\ref{FourierchiAB}), 
we relate the density correlator $\chi_{\rho\rho}$ to the the \emph{longitudinal} current correlator [see Eq.~(\ref{Nconserv})].
The latter is related to the viscosity as shown in Eq.~(\ref{KuboLR}),
namely $\zeta(\omega) + 4 \eta(\omega)/3 = \lim_{q\to 0} {m^2 \omega}\mathrm{Im}\chi_L(\bq,\omega)/{q^2}$.
We thus obtain
\bea 
\zeta(\omega) + {4\eta(\omega)}/{3} = \lim_{q\to 0}{\omega^3}\mathrm{Im}\chi_{\rho\rho}(\bq,\omega)/{q^4}.
\label{KuboLR2}
\eea
We discuss two situations where the contribution of $\zeta(\omega)$ vanishes and we can obtain
interesting results connecting $\eta(\omega)$ and density correlations. 

First, we focus on the unitary Fermi gas where $\zeta(\omega)$ vanishes at all $\omega$ (as shown in
Section \ref{Crossoversec}) and Eq.~(\ref{KuboLR2}) simplifies to Eq.~(\ref{bragg}).
Thus, the frequency-dependent shear viscosity $\eta(\omega)$ in a unitary Fermi gas can in principle be measured 
using an experiment like Bragg scattering, which directly probes $\mathrm{Im}\chi_{\rho\rho}$.

Second, let us look at the high-frequency regime $\epsilon_F \ll \omega \lesssim 1/(mr_0^2)$.
The $\zeta$ sum rule in Eq.~(\ref{zetasumrule2}) is convergent in the $r_0\to 0$ limit,
and thus $\zeta(\omega)$ must decay faster than $1/\omega$, while $\eta(\omega) \sim 1/\sqrt{\omega}$ 
(see Section \ref{highfrequencytailsec}). Thus, as $\omega \to \infty$, the bulk viscosity $\zeta(\omega)$
is much smaller than the shear viscosity $\eta(\omega)$ for all $1/(k_F a)$ and all $T/\epsilon_F$. 
Using Eq.~(\ref{KuboLR2}) we thus find
\bea 
\eta(\omega \to \infty) &\simeq& \lim_{\omega\to\infty}\lim_{q\to 0}\frac{3\omega^3}{4q^4}\mathrm{Im}\chi_{\rho\rho}(\bq,\omega)
\nonumber\\ &=&
\lim_{\omega\to\infty}\lim_{q\to 0}\frac{3\pi\omega^3}{4q^4}S(\bq,\omega),
\label{etahighomegaSqw}
\eea
The dynamic structure factor $S(\bq,\omega)$ is related to $\mathrm{Im}\chi_{\rho\rho}$ via the
fluctuation-dissipation theorem 
\bea
S(\bq,\omega) = \frac{\mathrm{Im}\chi_{\rho\rho}(\bq,\omega)}{\pi[1-\exp(-\beta\omega)]}.
\eea 
Our final result, given by Eq.~(\ref{sqomegatail}), for the high-frequency tail of $S(\bq,\omega)$ is obtained
by using the high-frequency tail, Eq.~(\ref{etahighomega}), of $\eta(\omega)$ in Eq.~(\ref{etahighomegaSqw}).

The high-frequency $\omega^{-7/2}$ tail of the dynamic structure factor result is a universal feature of short-range two-body interactions.
Remarkably, such a tail was first noticed in deep inelastic neutron scattering studies of superfluid $^4$He~\cite{Wong77}
and was subsequently understood in terms of hard-sphere gases~\cite{Kirkpatrick84}.
The high-frequency neutron scattering experiments probe the 
short distance properties of the two-body pair distribution function.
[In dilute Fermi gases, this is directly related to the contact $C$; see Eq.~(\ref{rho2})].    
It may seem surprising that such anomalous high-frequency tails arise even in dense systems like $^4$He.  
Recall that this behavior should be visible in a frequency range $n^{2/3}/m < \omega  < 1/mr^2_0$
in which the interaction ``looks" short-range.  Even in $^4$He, where $nr^3_0\lesssim 1$, such a frequency range can be found using deep inelastic neutron scattering, although the range is obviously much smaller than in dilute gases with $nr^3_0\ll 1$. 


\section{Comparison with sum rules for Relativistic Field Theories}
\label{comparisonsec}

There has been a considerable effort in the high-energy literature to understand the 
properties of viscosity spectral functions and their sum rules; see, e.g., Refs.~\cite{Teaney06,Romatschke09,Moore08}).  
In addition to understanding the transport coefficients within the AdS/CFT framework, 
this work seems to be motivated in part by an interest in reliably extracting transport coefficients 
of the quark-gluon plasma from lattice QCD calculations of Euclidean correlation functions. 
We briefly discuss here some similarities and differences between the results for
relativistic quantum field theories and those derived in this paper for
non-relativistic Fermi gases: Eqs.~(\ref{zetasumrule2A}) and (\ref{etasumrule3A}). 

There exist a number of Boltzmann calculations of the viscosity spectral functions in 
weak coupling QCD~\cite{Teaney06,Schafer09}. 
For the shear viscosity, the authors of Ref.~\cite{Schafer09} find the shear viscosity sum rule
\bea 
\frac{1}{\pi}\int^{\omega_c}_0d\omega \eta(\omega) = \frac{\varepsilon+P}{5},
\label{QCDshearsumrule}
\eea
where $g^4T\ll \omega_c \ll g^2T$ is a cutoff that removes a diverging contribution from a high-frequency tail.  
For the ${\cal{N}}=4$ supersymmetric Yang-Mills theory (SYM), Romatschke and Son~\cite{Romatschke09}
derived the following shear viscosity sum rule:
\bea 
\frac{1}{\pi}\int^{\infty}_0 d\omega [\eta(\omega)-\eta_{T=0}(\omega)] = \frac{\varepsilon}{5}.
\label{SUSY}
\eea
Here, a diverging vacuum contribution from a $T$-independent high-frequency tail has been subtracted out.  
We note that our $\eta$ sum rule in Eq.~(\ref{etasumrule3A}), though similar in structure, has one key difference.
The high-frequency tail for the Fermi gas is in general $T$-\emph{dependent}, because its coefficient is
set by the contact $C = k_F^4 {\cal C}[1/(k_F a), T/\epsilon_F]$. 

A non-perturbative calculation of the bulk viscosity sum rule in ${\cal{N}}=4$ supersymmteric Yang-Mills theory and pure Yang-Mills theory (QCD with no quarks) has been given recently by Romatschke and Son~\cite{Romatschke09}:
\bea 
\lefteqn{\frac{1}{\pi}\int^{\infty}_0 d\omega [\zeta(\omega)-\zeta_{T=0}(\omega)] =}&&\nonumber\\&&(3\varepsilon+P)(1-3c^2)- 4(\varepsilon-3P),
\label{QCDbulksumrule}
\eea
where $c\equiv \sqrt{\partial P/\partial \varepsilon}$ is the sound speed in relativistic hydrodynamics (with the speed of light equal to unity)~\cite{LLFM}. There are some differences and one very
interesting similarity with our $\zeta$ sum rule in Eq.~(\ref{zetasumrule2A}).  In contrast to the Fermi gas spectral function $\zeta(\omega)$, there is a need
to subtract out a divergent tail in Eq.~\ref{QCDbulksumrule} and this tail appears to be $T$-independent.
The interesting similarity is that in the ``conformal limit" $P = \varepsilon/3$, the right hand
side of Eq.~(\ref{QCDbulksumrule}) vanishes, analogous to the unitary Fermi gas.   

\section{Conclusions}
\label{conclusionssec}

In this paper we have derived various exact, non-perturbative results for the shear and bulk viscosities
of non-relativistic quantum fluids, focusing on the strongly interacting Fermi gas. Our main results were already 
summarized in the Introduction. To conclude, we discuss some open questions and how our results relate to them.

Most calculations~\cite{Bruun05,Schafer07} of the viscosity in strongly interacting Fermi gases
have so far been restricted to solving Boltzmann equations or using diagrammatic perturbation theory, in essence making a
quasiparticle approximation. Such results are valid in the high and low temperature regimes, but
not in the most interesting regime near and above $T_c$ where a quasiparticle approximation is questionable 
and the shear viscosity is known to be the smallest. It was recognized some time back~\cite{Randeria-reviews}
that there is a breakdown of Fermi liquid theory in the normal (i.e., non-superfluid) state of the
strongly interacting regime of the BCS-BEC crossover. It was shown that precursor pairing correlations
lead to a pseudogap \cite{Randeria-reviews}, which is a strong suppression of low-energy spectral weight
in various response functions. It is likely that no sharp quasiparticle excitations exist in this regime
near unitarity and just above $T_c$, but controlled calculations of dynamic quantities are very difficult. 

Quantum Monte Carlo methods have played an important role in determining the equilibrium thermodynamic properties 
of the unitary Fermi gas. However, results for transport coefficients are much less common, since they require
analytic continuation of imaginary time (Euclidean) data to the real axis~\cite{Aarts02}.
The sum rules we derive could serve as useful constraints on similar calculations for
strongly interacting Fermi gases.   

From an experimental point of view, the (static) shear viscosity for strongly interacting Fermi gases
has been estimated from studies of the damping of collective oscillations~\cite{Turlapov08}. We have shown
above that, at unitarity, the full frequency dependence of the shear viscosity spectral function
$\eta(\omega)$ can be obtained from two-photon Bragg spectroscopy. While it would be a challenging experiment (the density response being very small for small-$q$), this would give extremely important insights
into the strongly interacting Fermi gas, analogous to optical conductivity measurements of solids.

Finally, we return to the conjectured bound~\cite{Son-review} on the shear viscosity, Eq.~(\ref{bound}).
Proving or disproving the existence of a bound~\cite{bound} for non-relativistic quantum fluids like the strongly interacting Fermi gas
remains a challenging open problem. We hope that the spectral functions and sum rules derived here
constitute a step in this direction, just as they have for other well known inequalities
in quantum many-body physics.

\begin{acknowledgments}
ET would like to thank Shizhong Zhang, Georg Bruun, Joaqu\'in Drut, Vijay Shenoy, and Jason Ho for stimulating discussions.
MR would like to thank the participants at the International Conference on Recent Progress in Many Body Theories 
(RPMBT15) last summer for spurring his interest in this problem.   We thank Eric Braaten, Dam Son, and Sandip Trivedi for comments on the manuscript and Stefano Giorgini for sharing with us the Monte-Carlo data of Ref.~\cite{Astrakharchik04}.  
We gratefully acknowledge support from NSF-DMR 0706203, ARO W911NF-08-1-0338, and NSF-DMR 0907366.   
\end{acknowledgments}

\appendix

\section{Modified stress correlator}
\label{KuboPi0}

In the main paper we discussed current correlator and stress correlator representations of the 
bulk and shear viscosities. Here, we describe a third correlation function using an
explicitly defined operator $\widehat{\Pi}^0_{\alpha\beta}$ which
has been used to calculate the static shear viscosity $\eta = \mathrm{Re}\; \eta(\omega = 0)$.
We note that $\widehat{\Pi}^0$, which does not include the diagonal terms of the full stress tensor, 
cannot be used to calculate the bulk viscosity.

Let us define
\bea 
\hat{\Pi}^0_{\alpha\beta}(\br) = \frac{1}{2m} \sum_{i=1}^N 
\left\{ \hat{\bp}^{i}_{\alpha} \hat{\bp}^{i}_{\beta} , \delta(\br - \hat{\br}_i) \right\} 
\label{Pi0def}
\eea
where $\hat{\bp}^{i}_{\alpha}$ is the $\alpha$-component of the momentum operator for the $i$-th particle.
We emphasize that this is only one piece -- the kinetic part -- of the full stress tensor operator,
and omits other terms, such as the pressure. It is independent of the interaction potential
unlike the full stress tensor.
However, since the expectation value of the ``off-diagonal" part of $\widehat{\Pi}^0$ is identical to the hydrodynamic 
stress tensor in Eq.~(\ref{stress}), we expect that we can use $\widehat{\Pi}^0$ to compute the shear viscosity, at least in the 
low frequency limit. 

We define the correlator $\chi^{xy,xy}_{\Pi^0}$ by choosing
$\hat{A}=\hat{\Pi}^0_{\alpha\beta}(\bq)$ and $\hat{B}=\hat{\Pi}^0_{\mu\nu}(-\bq)$ in 
Eq.~(\ref{FourierchiAB}),
where
\bea 
\hat{\Pi}^{0}_{\alpha\beta}({\bq}) = \frac{1}{4m}\sum_{\bk\sigma}\hat{c}^{\dagger}_{\bk\sigma}\hat{c}_{\bk+\bq\sigma}(2k_{\alpha}+q_{\alpha})(2k_{\beta}+q_{\beta}).
\label{Pi0}
\eea
We can write the analog of Eq.~(\ref{KuboPiT}) as
\bea
\eta_0(\omega) = \lim_{q\to 0}\mathrm{Im}\chi^{xy,xy}_{\Pi^0}/\omega.
\eea
This is the form used by several authors~\cite{Bruun05,Peshier05} as a starting point for 
diagrammatic approximations. 

The sum rule for the modified stress correlator, Eq.~(\ref{Pi0}), simply follows from
the Kramers-Kronig relation:
\bea   
\lefteqn{\frac{1}{\pi}\int^{\infty}_0 d\omega\lim_{q\to 0}  \frac{\mathrm{Im}\chi^{xy,xy}_{\Pi^0}(\bq,\omega)}{\omega}=}&&\nonumber\\&& \lim_{\omega\to 0}\lim_{q\to 0}\frac{1}{2}\mathrm{Re}\chi^{xy,xy}_{\Pi^0}(\bq,\omega).
\label{sumrulePi0}
\eea
Ironically, it is seems harder to explicitly evaluate the
right hand side here than it is to calculate the exact sum rule in Eq.~(\ref{etasumrulePi}),
despite the simple operator $\widehat{\Pi}^0$ involved. The point is that
$\hat{\Pi}^0$ does not satisfy the Euler equation, Eq.~(\ref{euler-op}), and hence we  cannot 
relate it to the current.  
Thus, in contrast to the sum rules given by Eqs.~(\ref{etasumrule}) and (\ref{zetasumrule}) which involve the 
first frequency moment of the current correlator, we must 
deal directly with an inverse frequency moment in Eq.~(\ref{sumrulePi0}).  
Such an inverse moment is a generalized ``static susceptibility'', about which we do not seem to
know much, at least in this case. Unlike positive moment sum rules, it cannot be written in terms of commutators.


\section{Hydrodynamics}
\label{hydrosec}

In this Appendix, we review well-known hydrodynamic results for the current correlation functions~\cite{Kadanoff63,Hohenberg65}.  To keep the discussion as general as possible, we will use the full two-fluid hydrodynamic correlation functions that result from solving the linearized equations of two-fluid hydrodynamics~\cite{LLFM,ZNGbook}.  As written below, these correlation functions describe any superfluid with a two-component order parameter, including dilute two-component Fermi gases (see Ref.~\cite{Taylor09} and references therein), and reduce to standard hydrodynamic expressions in the normal phase above $T_c$.  
We start by writing down a relation between the longitudinal current correlation function $\chi_L(\bq,\omega)$ and the (mass) density response function $\chi_{\rho\rho}(\bq,\omega)$.  
(Recall that our current correlation function is the \textit{number} current correlation function and not the mass current correlation function generally used in the older literature~\cite{Kadanoff63,Hohenberg65}.  We will find it convenient in the analysis below to use the correlation function $\chi_{\rho\rho}$ for the mass density $\rho = mn$, however.)  Analogous to the result given by Eq.~(\ref{Correlationrelation}), the continuity equation $\partial_t n+ \bnab\cdot\bj=0$ can be used to find
\bea 
\omega^2\chi_{\rho\rho}(\bq,\omega) &=& (mq)^2\chi_L(\bq,\omega) - \bq\cdot\langle[\hat{\bj}_{\bq},\hat{\rho}_{-\bq}]\rangle\nonumber\\ &=&(mq)^2\chi_L(\bq,\omega) - \rho q^2.
\label{Nconserv}
\eea
We can now use the hydrodynamic expression for $\chi_{\rho\rho}(\bq,\omega)$ to obtain an explicit hydrodynamic expression for $\chi_L(\bq,\omega)$.  In the hydrodynamic regime, the density response function is (see, e.g., Eq.~(4.32) in Ref.~\cite{Hohenberg65}):
\bea 
\lefteqn{\frac{\chi_{\rho\rho}(\bq,\omega)}{\rho q^2} =}&&\nonumber\\&&\!\!\!\!\!\!\!\! 
\frac{-\omega^2 + q^2\frac{\rho_s T s^2}{\rho_n c_v} - iq^2\omega\Gamma}
{(\omega^2 - u^2_1q^2 + iD_1q^2\omega)(\omega^2-u^2_2q^2 + iD_2q^2\omega)}.
\label{chirhorho}
\eea
Here, $u_1$ and $u_2$ are the speeds of first and second sound, respectively.  They can be shown to satisfy the following identity (see, e.g., Eq.~(14.39) in Ref.~\cite{ZNGbook}):
\bea u^2_1 + u^2_2 = \frac{\rho_s T s^2}{\rho_n c_v} + c^2_s \ \ \ {\rm with} \ \ \ 
c^2_s = \left({\partial P}/{\partial \rho}\right)_{s}.
\label{u1u2identity}\eea
Recall that $s\equiv S/N$ is the entropy per particle.  $c_v\equiv T(\partial s/\partial T)_{\rho}$ is the specific heat per unit mass at constant volume.  $\rho_s$ and $\rho_n$ are the superfluid and normal fluid densities.  The damping coefficients $\Gamma$, $D_1$, and $D_2$ obey the following identities:
\bea \Gamma \equiv \frac{\kappa}{\rho c_v} + \frac{\rho_s}{\rho_n}\left(\frac{4\eta/3+\zeta_2}{\rho}-2\zeta_1+\rho\zeta_3\right)\label{D02}\eea
and
\bea D_1+D_2 = \Gamma + \frac{1}{\rho}\left(4\eta/3+\zeta_2\right).\eea
Here, $\kappa$ is the thermal conductivity and $\zeta_1$, $\zeta_2$, and $\zeta_3$ are the bulk viscosities associated with the different types of motion that can arise in the superfluid phase~\cite{LLFM}.  Above $T_c$, $\zeta_2$ reduces to the usual bulk viscosity $\zeta$ and the remaining bulk viscosities do not contribute.  

After some straightforward but lengthy algebra, one can show from Eqs.~(\ref{Nconserv}) and (\ref{chirhorho}) that
\bea 
\lim_{\omega\to 0}\lim_{q\to 0}\frac{m^2\omega^2}{2q^2}\mathrm{Re}\chi_{L}(\bq,\omega)\! = \! - 
\frac{\rho}{2}\left[u^2_1 + u^2_2-\frac{\rho_s T s^2}{\rho_n c_v}\right].\nonumber\\
\eea
Using Eq.~(\ref{u1u2identity}) in this expression gives the result in Eq.~(\ref{difference2}).  

The transverse current correlation function is given by (see Eq.~(4.49) in Ref.~\cite{Hohenberg65})
\bea 
m^2\chi_T(\bq,\omega) = \frac{\eta q^2}{\eta q^2/\rho_n - i\omega}.
\label{chiT}
\eea  
From Eq.~(\ref{chiT}), we see that the real part of the transverse current correlation function is proportional to $q^4$, leading to the result in Eq.~(\ref{difference2}).  


\section{Contact} 
\label{ContactAppendix}

This Appendix consists of two parts. In the first part, 
we briefly recall, for completeness, some basic properties of
Tan's contact; more details may be found in the original
references~\cite{Tan08,Braaten08}. In the second part,
we use the same techniques, within a real-space formulation~\cite{Zhang08}, 
to derive Eqs.~(\ref{identity1}) and (\ref{identity2})
for $\overline{V}'$ and $\overline{V}''$.

The contact $C$ can be defined by the large-$k$ tail of the momentum distribution function
$\lim_{k\to \infty} n_{\bk} = {C}/{k^4}$. This leads to a kinetic energy density
$\langle \hat{K} \rangle = 2\int d^3\bk \; (k^2/2m) \; n_{\bk}$ with a linearly divergent
piece that goes like $C\Lambda/(2\pi^2m)$, where $\Lambda \equiv 1/r_0$ is the ultraviolet cutoff.
The potential energy density is given by~\cite{Braaten08}
\bea 
\langle \hat{V} \rangle  = \frac{ C}{4\pi ma} - \frac{ \Lambda C}{2\pi^2 m},
\label{eint2}
\eea
so that the total energy density $\varepsilon = \langle \hat{K} \rangle  + \langle \hat{V} \rangle$
is finite in the $\Lambda = 1/r_0 \to \infty$ limit. We will freely use these
results, together with those obtained from the short-distance
properties of two-body density matrix, to evaluate the quantities of interest
for our sum rules.

At short distances, $r_0\lesssim r \ll k^{-1}_F$, the two-body density matrix for a 
two-component dilute Fermi gas has the structure~\cite{Zhang08}
\bea 
\int\!\! d^3\bR \Big\langle \hat{\psi}^{\dagger}_{\uparrow}(\bR\!+\!\frac{\br}{2})
\hat{\psi}^{\dagger}_{\downarrow}(\bR\!-\!\frac{\br}{2})
\hat{\psi}_{\downarrow}(\bR\!-\!\frac{\br}{2})\hat{\psi}_{\uparrow}(\bR\!+\!\frac{\br}{2})\Big\rangle
\nonumber\\
\equiv {\cal F}(\br) = \frac{C}{16\pi^2}\left(\frac{1}{r}-\frac{1}{a}\right)^2.
\label{rho2-again}
\eea 
Let us use this to compute the interaction energy density
\bea 
\langle \hat{V} \rangle =  \frac{C}{4\pi}\int dr V(r)\left(1 - {r}/{a}\right)^2.
\label{moment0b2}\eea
It is easy to see that for $r_0\to 0$, we may drop the $(r/a)^2$ term
in the integrand as it gives a vanishingly small contribution. 
Using
\bea 
X_n = \frac{C}{4\pi} \int dr V(r) (r/a)^n \ \ \ \ (n=0,1)
\label{xdef}
\eea 
we thus obtain
\bea 
\langle \hat{V} \rangle = X_0 - 2 X_1.
\label{moment0c}
\eea
Similarly, Eqs.~(\ref{moment1b}) and (\ref{moment2b}) may be written
as 
\bea
\overline{V}' = -X_0 + 4 X_1
\label{moment1c}
\eea
and
\bea \overline{V}'' =2 X_0 - 12 X_1.
\label{moment2c}
\eea

Next, we wish to determine the integrals $X_0$ and $X_1$
in the limit where the range of the potential vanishes: $r_0 \to 0$. 
Comparing the results given by Eqs.~(\ref{eint2}) and (\ref{moment0c}) for $\langle \hat{V} \rangle$,
it is evident that $X_0$ is linearly divergent in $\Lambda$ and $X_1$ is finite
as $\Lambda = 1/r_0 \to \infty$.
But there is no way to determine the finite part of $X_0$ from this comparison alone.
We need one additional piece of information to determine
$X_0$ and $X_1$. We get this from the relation
\bea 
P = {2}\varepsilon/3 + \langle \hat{V}\rangle/{3} + \overline{V}'/{3}.
\label{P0}
\eea
for the pressure $P$ which is derived in Appendix \ref{Pressuresec}.
Using Eqs.~(\ref{moment0c}) and (\ref{moment1c}), we see
that the divergent terms ($X_0$) cancel and the pressure
\bea 
P = {2}\varepsilon/{3} + {2}X_1/3
\label{P1}
\eea
is finite as $r_0 \to 0$. Comparing this with \cite{Tan08}
$P = 2\varepsilon/3 + {C}/(12\pi m a)$ we obtain
\bea 
X_1 = {C}/(8 \pi m a).
\label{x1}
\eea
Substituting this into Eq.~(\ref{moment0c}) for
$\langle \hat{V} \rangle$, and comparing with 
Eq.~(\ref{eint2}), we find
\bea 
X_0 = - C \Lambda /(2 \pi^2 m) - {C}/(2 \pi m a).
\label{x0}
\eea
Using these results for $X_0$ and $X_1$ in 
Eqs.~(\ref{moment1c}) and (\ref{moment2c})
we obtain Eqs.~(\ref{identity1}) and (\ref{identity2})
for $\overline{V}'$ and $\overline{V}''$, respectively.


\section{Pressure} 
\label{Pressuresec}

In the main text of this paper, we have suppressed factors of the volume $\Omega$ at all places by setting it equal to unity.
In this Appendix, we re-introduce factors of $\Omega$ in order to use
the thermodynamic relation $P = -(\partial F/\partial \Omega)_{T,N}$ to derive a microscopic expression for the pressure.
To evaluate this, we use the Feynman-Hellmann formula 
$(\partial F/\partial\lambda)_{T,N} = \langle \partial \hat{H}/\partial \lambda\rangle$ 
treating the volume $\Omega$ as a parameter $\lambda$ in the Hamiltonian
\bea \hat{H} = \sum_{\bk\sigma}\varepsilon_{\bk}\hat{c}^{\dagger}_{\bk\sigma}\hat{c}_{\bk\sigma}\! +\! \frac{1}{\Omega}\!\sum_{\bk\bk'\bp}V(p)\hat{c}^{\dagger}_{\bk+\bp\uparrow}\hat{c}^{\dagger}_{\bk'-\bp\downarrow}\hat{c}_{\bk'\downarrow}\hat{c}_{\bk\uparrow}.\label{HV}\eea
The volume enters in two ways:
(i) explicitly, though the $\Omega^{-1}$ factor in front of the interaction term, and
(ii) implicitly, through the discrete wavevectors $\bk = \left(2\pi/\Omega^{1/3}\right)(n_x,n_y,n_z)$.  One can replace the momentum sums
by sums over the discrete indices $n_\alpha = 0,1,...$, and the operators only depend on these indices.  
Thus, in addition to the explicit $\Omega^{-1}$ factor, the kinetic energy $\varepsilon_{\bk}=\bk^2/2m$ and the two-body potential $V(p)$ also
depend on the volume $\Omega$.
  
Using $3\Omega \left[\partial F(\bk)/\partial\Omega\right] = - k_\alpha \left[\partial F(\bk)/\partial k_{\alpha}\right]$ 
(with the summation convention) and the definition of $\overline{V}'$ given in Eq.~(\ref{Vpp}),
we find the result 
\bea 
P = \frac{2}{3}\varepsilon + \frac{1}{3}\langle \hat{V}\rangle + 
\frac{1}{3}\overline{V}'.
\label{P2}\eea


\section{Thermodynamics of the BCS-BEC Crossover}
\label{universalthermosec}

In this Appendix, we simplify the form of the bulk viscosity sum rule in Eq.~(\ref{zetasumrule2})
and derive the result given by Eq.~(\ref{zetasumrule2A}) using thermodynamic scaling arguments~\cite{Ho04} and the
Tan relations~\cite{Tan08}. We begin by writing the energy density of a two-component Fermi gas in the scaling form
\bea
\varepsilon = n \epsilon_F {\cal{E}}(x,s),
\label{energy}
\eea
which is valid across the entire BCS-BEC crossover. Here, ${\cal{E}}$ is a dimensionless function of the interaction
parameter $x = 1/(k_F a)$ and the entropy per particle $s\equiv S/N$. We find it convenient to use 
$s$, rather then more familiar variable $T/\epsilon_F$, because we will need to
evaluate adiabatic derivatives below. The density dependence of the Fermi energy 
and Fermi wavevector are given by $\epsilon_F = (3\pi^2 n)^{2/3}/2m$ and
$k_F = (3\pi^2 n)^{1/3}$ respectively, and $a$ is the $s$-wave scattering length.

We first calculate the adiabatic sound speed which enters the right hand side of the 
$\zeta$ sum rule in Eq.~(\ref{zetasumrule2}). Using the definition
$c_s^2 = (\partial P / \partial \rho)_s$, where $\rho=mn$, together with 
Tan's pressure relation, Eq.~(\ref{tanP}), we find
\bea
\frac{\rho c_s^2}{2} = \frac{n}{2} \left[\frac{2}{3}\left(\frac{\partial \varepsilon}{\partial n}\right)_{s}
+ \frac{1}{12\pi m a} \left(\frac{\partial C}{\partial n}\right)_{s} \right].
\label{rhoc2}
\eea
The derivatives at constant $s$ are evaluated as follows.
The first term is
$({\partial \varepsilon}/{\partial n})_{s} = (5\varepsilon/3n) - (\epsilon_F x/3)({\partial {\cal{E}}}/{\partial x})_{s}$.
We compute $({\partial {\cal{E}}}/{\partial x})_{s}$ using Tan's adiabatic relation, Eq.~(\ref{tanA}), and obtain
$n({\partial \varepsilon}/{\partial n})_{s} = 5\varepsilon/3 - C/(12\pi m a)$.
To calculate the second term in Eq.~(\ref{rhoc2}), we rewrite the contact in the scaling form
\bea
C = k^4_F \widetilde{C}(x,s)
\eea
where $\widetilde{C}$ is a dimensionless function on its arguments.
After some simple algebra, we find 
$n({\partial C}/{\partial n})_{s} = 4C/3 - (3 a)^{-1}({\partial C}/{\partial a^{-1}})_{s}$.
Adding up all of the contributions to the bulk viscosity sum rule, we find
\bea 
\frac{5\varepsilon}{9} + \frac{C}{12 \pi m a} - \frac{\rho c_s^2}{2} = 
\frac{1}{72 \pi m a^2} \left( \frac{\partial C}{\partial a^{-1}} \right)_s.
\label{zetasumrule-simple}
\eea   

That the $\zeta$ sum rule vanishes at unitarity can also be seen directly from thermodynamic scaling arguments, without introducing the contact.  Using Eq.~(\ref{energy}), we see that $P=-(\partial (\varepsilon \Omega)/\partial \Omega)_{S,N}=2\varepsilon/3$ at unitarity [as anticipated by Eq.~(\ref{tanP})]~\cite{Thomas05}. Using this, we also find that the adiabatic sound speed at all temperatures is given by $c^2_s = (1/m)\left(\partial P/\partial n\right)_{s} = 5P/3\rho$.  
Combining these results, one immediately obtains the result in Eq.~(\ref{ZetaSumUnitarity}) that the bulk viscosity sum rule vanishes there.


\begin{thebibliography}{99}
\bibitem{wallnote} A Galilean shift such as that used to arrive at Eq.~(\ref{Hshift}) is strictly only valid for a uniform velocity field $\bu(\br,t)=\bu(t)$.  In this case, Eq.~(\ref{Hshift}) gives the Hamiltonian in the rest frame of the ``walls" bounding the system in accordance with the principle of Galilean relativity~\cite{Baymbook}.  However, the fact that the velocity field is (necessarily) nonuniform means that translational symmetry is broken and there is no suitable reference frame in which the system can be in equilibrium.  Nonetheless, in the long wavelength limit required to define the viscosity coefficients, Eq.~(\ref{Hshift}) reduces to the Hamiltonian in the rest frame of the walls, meaning that it is suitable for defining equilibrium thermodynamic averages.
\bibitem{limits}
Consider, e.g., the induced current in the $\alpha=x$ direction, and set $q_z = 0$.
The transverse correlator is obtained by first taking the $q_x\to 0$, 
followed by $q_y\to 0$ limit. 
Interchanging the order, one finds the longitudinal result.
\bibitem{eta-zero} 
We argue that $(2\alpha_b + \beta_b - 2) > 0$.  Otherwise, $\eta$ would have potentially divergent contributions.
As discussed above, $\eta$ is analogous to the resistivity and does not have a $\delta(\omega)$ piece.
This is very different from superconductors which have a $\delta(\omega)$ term in their frequency-dependent conductivity.
\bibitem{uv-cutoff}
Recall that $\Lambda = 1/r_0$ in Eq.~(\ref{eint}) comes from an ultraviolet cutoff imposed on the two-body $T$-matrix.  
In the center-of-mass frame of the two-body scattering problem, this is equivalent to imposing an energy cutoff 
$\Lambda^2/2m_{\mathrm{r}}$ with reduced mass $m_{\mathrm{r}} = m/2$; hence the cutoff in Eq.~(\ref{etasumrule3}).
\bibitem{bound} For relativistic systems without a conserved density, $\eta/s$ is the natural quantity to look at.  For non-relativistic systems, a more natural quantity to look at would be $\eta/\rho_n$, where $\rho_n$ is the normal fluid density defined in Sec.~\ref{rhon-section}.
\end{thebibliography}
\end{document}